\documentclass[]{aa}
\usepackage{graphicx}
\usepackage{txfonts}
\usepackage{natbib}
\bibpunct{(}{)}{;}{a}{}{,}

\begin{document}

   \title{The UV spectrum of HS~1700+6416}

   \subtitle{II. FUSE observations of the He\,II Lyman alpha forest}

   \author{C. Fechner\inst{1}
          \and
          D. Reimers\inst{1}
          \and
          G.~A. Kriss\inst{2}
          \and
	  R. Baade\inst{1}
	  \and
	  W.~P. Blair\inst{3}
	  \and
	  M.~L. Giroux\inst{4}
	  \and
	  R.~F. Green\inst{5}
          \and
	  H.~W. Moos\inst{3}
	  \and
	  D.~C. Morton\inst{6}
	  \and
	  J.~E. Scott\inst{2,}\inst{7}
          \and
          J.~M. Shull\inst{8}
          \and
	  R. Simcoe\inst{9}
          \and
	  A. Songaila\inst{10}
	  \and
	  W. Zheng\inst{3}
          }

   \offprints{C. Fechner}

   \institute{Hamburger Sternwarte, Universit\"at Hamburg,
              Gojenbergsweg 112, 21029 Hamburg, Germany,\\
              \email{[cfechner,dreimers,rbaade]@hs.uni-hamburg.de}
         \and
             Space Telescope Science Institute, 3700 San Martin Drive,
	     Baltimore, MD 21218, USA,
             \email{gak@stsci.edu} 
	 \and
	     Department of Physics \& Astronomy, The Johns Hopkins University,
	     3400 North Charles Street, Baltimore, MD 21218, USA
	 \and
             Dept. of Physics, Astronomy, and Geology,
	     East Tennessee State University, Johnson City, TN 37614, USA
	 \and
	     Large Binocular Telescope Observatory,
	     933 N. Cherry, Tucson, AZ  85721, USA
	 \and
	     Herzberg Institute of Astrophysics, 
	     National Research Council of Canada, 5071 West Saanich Road, 
	     Victoria, BC V9E 2E7, Canada
	 \and
             present address: 
	     Department of Physics, Astronomy, and Geosciences, 
	     Towson University, Towson, Maryland 21252, USA
         \and
             CASA, Department of Astrophysical and Planetary Sciences,
	     University of Colorado, Boulder, CO 80309, USA
	 \and
	     MIT Center for Space Research, 
	     77 Massachusetts Ave. 37-664B, Cambridge, MA 02139, USA
	 \and
             Institute for Astronomy, University of Hawaii,
	     2680 Woodlawn Drive, Honolulu, HI 96822, USA
             }

   \date{Received February 1, 2006; accepted April 28, 2006}

   \abstract{We present the far-UV spectrum of the quasar \object{HS~1700+6416} taken with FUSE.
This QSO provides the second line of sight with the \ion{He}{ii} absorption resolved into a Ly$\alpha$ forest structure.
Since HS~1700+6416 is slightly less redshifted ($z_{\mathrm{em}} = 2.72$) than \object{HE~2347-4342}, we only probe the post-reionization phase of \ion{He}{ii}, seen in the evolution of the \ion{He}{ii} opacity, which is consistent with a simple power law.
The \ion{He}{ii}/\ion{H}{i} ratio $\eta$ is estimated using a line profile-fitting procedure and an apparent optical depth approach, respectively.
The expected metal line absorption in the far-UV is taken into account as well as molecular absorption of galactic H$_2$. 
About 27\,\% of the $\eta$ values are affected by metal line absorption.
In order to investigate the applicability of the analysis methods, we create simple artificial spectra based on the statistical properties of the \ion{H}{i} Ly$\alpha$ forest.
The analysis of the artificial data demonstrates that the apparent optical depth method as well as the line profile-fitting procedure lead to confident results for restricted data samples only ($0.01 \le \tau_{\ion{H}{i}} \le 0.1$ and $12.0 \le \log N_{\ion{H}{i}} \le 13.0$, respectively).
The reasons are saturation in the case of the apparent optical depth and thermal line widths in the case of the profile fits.
Furthermore, applying the methods to the unrestricted data set may mimic a correlation between the $\ion{He}{ii}/\ion{H}{i}$ ratio and the strength of the \ion{H}{i} absorption.
For the restricted data samples a scatter of $10 - 15$\,\% in $\eta$ would be expected even if the underlying value is constant.
The observed scatter is significantly larger than expected, indicating that the intergalactic radiation background is indeed fluctuating.
In the redshift range $2.58 < z < 2.72$, where the data quality is best, we find $\eta \sim 100$, suggesting a contribution of soft sources like galaxies to the UV background. 

}

   \maketitle

%________________________________________________________________

%******************************************************************************
\section{Introduction}

The shape and the strength of the intergalactic UV background plays an important role in governing the evolution of the ionization state of the intergalactic medium (IGM).
Since the IGM is observed to be highly ionized, ionization corrections have to be applied to measure properties of interest like its metallicity.
The strength of the UV background, at least at the \ion{H}{i} ionization edge, can be derived from the proximity effect \citep[e.g.][]{scottetal2000}, whereas more information is needed to constrain the shape of the ionizing radiation field.
\citet{agafonovaetal2005} introduced a method to estimate the shape of the ionizing background from the observation of metal line absorbers.

Theoretical calculations model the shape of the UV background as the radiation of sources attenuated while propagating though the IGM.
Quasars have been adopted as sources \citep{haardtmadau1996, fardaletal1998} whose radiation has been filtered by the IGM.
Since hydrogen and helium are the most abundant elements in the IGM, the resulting shape of the mean UV background is characterized by a sharp break at 1\,Ryd and another break at 4\,Ryd.

Observational results, however, suggest that the intergalactic UV background is dominated not only by the filtered radiation of quasars but also by a significant contribution from the softer radiation of galaxies.
This evidence comes from the study of the \ion{H}{i} opacity \citep[e.g.][]{kirkmanetal2005} or the number density evolution of \ion{H}{i} absorbers \citep[e.g.][]{zhangetal1997, bianchietal2001}, and also the derivation of reasonable metallicities from observed metal line absorption requires softer radiation \citep[][]{aguirreetal2004}.
Recent computations of the UV background therefore consider the contribution of galaxy radiation \citep[e.g.][]{haardtmadau2001}.
Furthermore, \citet{schaye2004} and \citet{miraldaescude2005} pointed out that local sources may dominate over the mean background.

Besides \ion{H}{i} opacity measurements and the investigation of metal line systems, \ion{He}{ii} provides another observational probe of the UV background.
Comparing the strength of the \ion{He}{ii} Ly$\alpha$ absorption to that of \ion{H}{i} leads to the column density ratio $\eta = N_{\ion{He}{ii}}/N_{\mathrm{\ion{H}{i}}}$, which is directly related to the ionizing background at the \ion{H}{i} and \ion{He}{ii} photoionization edge \citep[e.g.][]{fardaletal1998}.
Low values are expected for hard radiation, while high values should be measured if the ionizing radiation is dominated by soft sources.
However, the observation of the \ion{He}{ii} Ly$\alpha$ transition ($\lambda_{\mathrm{rest}} = 303.7822\,\mathrm{\AA}$) is very difficult.
An unabsorbed, UV-bright QSO is needed at redshift $z \gtrsim 2$ to be observable with HST/STIS or FUSE \citep{picardjakobsen1993}.
Up to now six of such quasars have been found \citep{jakobsenetal1994, davidsenetal1996, reimersetal1997, andersonetal1999, zhengetal2004b, reimersetal2005c}.
Comparisons of the \ion{He}{ii} opacity to the \ion{H}{i} opacity along a number of these sightlines point toward a dominant contribution to the ionizing UV background from hard sources, such as quasars, but with significant contributions also from soft sources such as star-forming galaxies \citep{heapetal2000, krissetal2001, smetteetal2002, zhengetal2004b, shulletal2004}.

The \ion{He}{ii} Ly$\alpha$ forest has been resolved for the first time towards the quasar HE~2347-4342 with FUSE \citep{krissetal2001, shulletal2004, zhengetal2004}.
Different analysis methods have been applied to derive the column density ratio (line profile fitting was used by \citet{krissetal2001} and \citet{zhengetal2004}, while \citet{shulletal2004} applied an apparent optical depth method) and lead to the same results:
The UV background is highly variable on small scales with a fluctuation of the column density ratio between $1 \lesssim \eta \lesssim 1000$ or even more.
Additionally, \citet{shulletal2004} find an apparent correlation between the $\eta$ value and the strength of the \ion{H}{i} absorption in the sense, that $\eta$ appears to be larger in voids.

In this work we present FUSE observations of the quasar HS~1700+6416, which provides the second line of sight of a resolved \ion{He}{ii} Ly$\alpha$ forest.
Owing to the variability of the UV flux of this QSO \citep{reimersetal2005b}, the data is of comparable quality like those of HE~2347-4342 ($S/N \sim 5$, $R \approx 20\,000$).
The emission line redshift of HS~1700+6416 is $z_{\mathrm{em}} = 2.72$, i.e.\ this line of sight probes the post-reionization universe, while in the spectrum of HE~2347-4342 the end of the \ion{He}{ii} reionization phase has been detected \citep{reimersetal1997}.
HS~1700+6416 has a rich metal line spectrum in the optical as well as in the UV comprising seven Lyman limit systems (LLS), which has been analyzed by several authors \citep[e.g.][]{reimersetal1992, vogelreimers1993, vogelreimers1995, koehleretal1996, petitjeanetal1996, trippetal1997, simcoeetal2002, simcoeetal2006}.
Thus, metal line absorption features are also expected to arise in the far-UV polluting the Ly$\alpha$ forest absorption.
Therefore, we will include a model of the metal line absorption in the analysis to obtain unbiased results.
For this purpose, we derived a prediction of the metal line spectrum in the far-UV \citep{fechneretal2005a}, which we will consider in this work.

This paper is organized as follows:
After presenting the observations in Sect.\ \ref{observations2} and describing the extrapolation of the continuum for the FUSE spectral range in Sect.\ \ref{continuumdef}, we present the far-UV metal line prediction derived in \citet{fechneretal2005a} in Sect.\ \ref{mls} and give a comparison to the data.
In Sect.\ \ref{simulations} we discuss our artificial data and analyze them using the line profile-fitting procedure and the apparent optical depth method to investigate the applicability of these methods.
The results derived from the observed data are presented and discussed in Sect.\ \ref{results}.
We conclude in Sect.\ \ref{conclusion}.

%******************************************************************************
\section{Observations}\label{observations2}

The Far Ultraviolet Spectroscopic Explorer (FUSE) uses four independent optical
channels to enable high-resolution spectroscopy in the far-ultraviolet wavelength range below $1200\,\mathrm{\AA}$.
In each channel a primary mirror gathers light for a Rowland-circle spectrograph. 
Two-dimensional photon-counting detectors record the dispersed spectra. 
Two of the optical channels use LiF coatings on the optics to cover the $990 - 1187\,\mathrm{\AA}$ wavelength range.
The other two channels cover shorter wavelengths down to $912\,\mathrm{\AA}$ using SiC-coated optics.  
\citet{moosetal2000} give a full description of FUSE, and \citet{sahnowetal2000} summarize its in-flight performance.

We observed HS~1700+6416 with FUSE through the 30\arcsec-square low-resolution apertures (LWRS) during four epochs in October 1999, May 2002, February/March 2003, and May 2003.
Table \ref{obs} gives details of the individual observations.
Between the exposures in May 2002 and 2003 the UV brightness of HS~1700+6416 increased by a factor $\sim 3$ \citep[see][]{reimersetal2005b}.
Therefore, the final data are of better quality than we originally expected.
The total exposure time is roughly $791\,118\,\mathrm{s}$ with $290\,230\,\mathrm{s}$ during orbital night.

\begin{table}
  \begin{center}
  \caption[]{Log of FUSE Observations of HS1700+64}
  \label{obs}
  \begin{tabular}{l c r r}
    \hline\hline
    \noalign{\smallskip}
    Observation ID & Date & $t_{\mathrm{total}}\,(\mathrm{s})$ & $t_{\mathrm{night}}\,(\mathrm{s})$ \\
    \noalign{\smallskip}
    \hline
    \noalign{\smallskip}
       P1100401 & 1999 Oct 14 &  57\,038 & 23\,275 \\
       C1230101 & 2002 May 15 &  54\,710 & 17\,702 \\
       C1230102 & 2002 May 16 &  53\,821 & 15\,663 \\
       C1230103 & 2003 Feb 27 & 156\,963 & 93\,828 \\
       C1230104 & 2003 Mar 03 & 130\,605 & 51\,584 \\
       P3060101 & 2003 May 02 &  61\,326 & 18\,520 \\
       P3060102 & 2003 May 03 &  57\,952 & 17\,863 \\
       P3060103 & 2003 May 10 &  91\,590 & 23\,824 \\
       Z0120101 & 2003 May 03 &  53\,896 & 14\,943 \\
       Z0120102 & 2003 May 04 &  43\,520 &  9\,702 \\
       Z0120103 & 2003 May 05 &  21\,064 &  3\,019 \\
       Z0120104 & 2003 May 10 &   9\,916 &    307 \\
    \noalign{\smallskip}
    \hline
  \end{tabular}
  \end{center}
\end{table}

We use the FUSE calibration pipeline CALFUSE V2.2.1 to process the data.
The earliest versions of the calibration pipeline are described by \citet{sahnowetal2000}. 
CALFUSE V2.2.1 is described in detail on the FUSE web site (http://fuse.pha.jhu.edu/analysis/pipeline\_reference.html).
The major improvements to the pipeline since the description of \citet{sahnowetal2000} that are relevant for our work are that data screening for bursts has been automated, and that a two-dimensional, multicomponent model for the background has been implemented.
The background model takes into account the uniform (but time-varying) particle
background in the detector as well as the spatially varying and time-varying portions of the background due to scattered light within the instrument.

Since we are working at the very limits of instrument sensitivity, we take additional non-standard steps in our processing to reduce the background. 
We restrict the data to orbital night only (scattered light is ten times brighter during orbital day), and we select only detector events with pulse heights in the range $4 - 16$. 
This reduces the background by an additional factor of two while still retaining photometric accuracy in the final result.
Finally, since the instrument throughput is roughly a factor of three higher in the LiF channels compared to the SiC channels, we only use data from the LiF channels. 
The final spectrum therefore only covers the wavelength range $1000 - 1180\,\mathrm{\AA}$, with a gap at $1080 - 1086\,\mathrm{\AA}$ due to the detector layout.

Each observation interval listed in Table \ref{obs} is processed separately.
To assure the best spectral resolution and an accurate wavelength scale, we cross-correlate the overlapping wavelength regions of each extracted spectrum with each other. 
No wavelength shifts were identified that exceeded our final bin size of $0.05\mathrm{\AA}$.
The extracted spectra from each detector segment and exposure are then merged onto a uniform wavelength scale in $0.05\mathrm{\AA}$ bins.
To maximize the resulting signal-to-noise ratio, each extracted spectrum is weighted by
the product of the exposure time and the continuum flux at $1000\,\mathrm{\AA}$.
This effectively weights each spectrum by the expected relative number of photons in each spectrum, which is the optimal weighting for data with Poisson-distributed errors.
The resulting spectrum shown in Fig.\ \ref{continuum} has $S/N \sim 5$ and a
spectral resolution of $R \sim 20\,000$.

In the final merged spectrum we notice some potential problems with the zero flux level, particularly in the data at wavelengths shortward of $1080\,\mathrm{\AA}$.
Strong, saturated \ion{H}{i} absorption is detected at $z \approx 2.315$ and $z \approx 2.433$ arising from LLS at these redshifts.
At the corresponding positions in the FUSE spectrum ($\sim 1007\,\mathrm{\AA}$ and $\sim 1043\,\mathrm{\AA}$) strong but slightly unsaturated features are detected.
Since the column density ratio $\eta$ should be $>1$ because of physical plausibility, these features actually have to be saturated.
However, the troughs of these presumably saturated lines are consistent with zero flux within the errors of our $S/N \sim 5$ spectrum.
We note the systematic errors that may result from this in our future discussion, but note that data at longer wavelengths ($\lambda > 1086\,\mathrm{\AA}$) does not appear to have the same problem.
Saturated \ion{H}{i} features at $z \sim 2.597$, $2.602$, $2.634$, and $2.683$ show also saturated \ion{He}{ii} counterparts at $\sim 1092\,\mathrm{\AA}$, $1094\,\mathrm{\AA}$, $1104\,\mathrm{\AA}$, and $1119\,\mathrm{\AA}$ in the FUSE spectrum (see e.g.\ Fig\ \ref{metals}).
Thus, we are confident we can obtain unbiased results in the range $2.58 < z < 2.72$.
Problems with the interpretation of the results due to inaccuracies in the zero flux level in the range $2.29 < z < 2.56$ will be discussed later.

The corresponding \ion{H}{i} Ly$\alpha$ forest was observed with Keck.
We have access to two datasets already published by \citet{songaila1998} and \citet{simcoeetal2002}.
These two datasets are co-added obtaining a resulting spectrum with a total exposure time of 84\,200 s and a signal-to-noise of $S/N \sim 100$ at $4000\,\mathrm{\AA}$.
The co-added spectrum covers the wavelength range $3680 - 5880\,\mathrm{\AA}$ with a resolution of $R \sim 38\,500$.
The Simcoe data covers wavelengths down to $\sim 3220\,\mathrm{\AA}$ with less
signal-to-noise.
We also have used the absorption lines identified in this portion of the spectrum to constrain the metal line system models \citep[see][]{fechneretal2005a}.

%******************************************************************************
\section{Continuum definition}\label{continuumdef}

\begin{figure*}
  \centering
  \resizebox{\hsize}{!}{\includegraphics[bb=320 35 555 785,angle=-90,clip=]{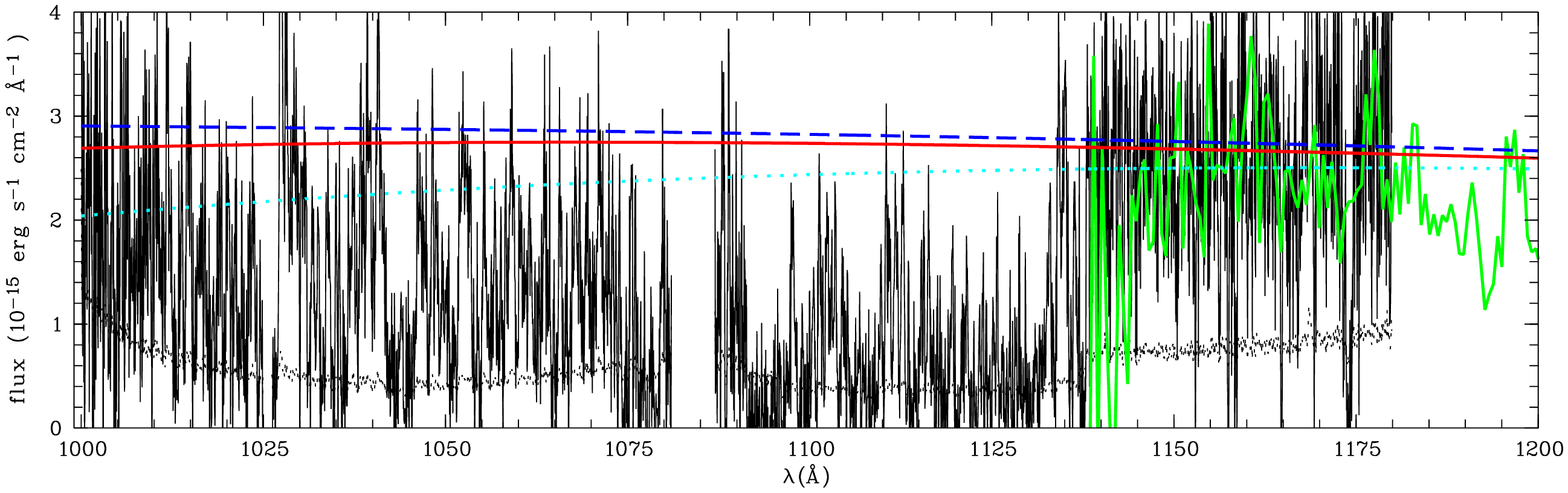}}
  \caption{The FUSE data and the extrapolated continuum. 
The solid line presents the continuum considering the \citet{cardellietal1989} extinction curve and $E(B-V) = 0.025$ \citep{schlegeletal1998}.
For comparison we show the extrapolated continuum using $E(B-V) = 0.060$ adopted from X-ray observations \citep{reimersetal1995} and the \citet[][dotted line]{cardellietal1989} or the \citet[][dashed line]{pei1992} extinction curve, respectively.
Furthermore, we overlay the low resolution STIS data which was taken simultaneously with the FUSE data in May 2003.
For a higher resolved presentation of the normalized FUSE spectrum see Fig. \ref{metals}.
  }
  \label{continuum}
\end{figure*}

In order to estimate the continuum in the FUSE portion of the spectrum we use low resolution data taken with HST/STIS in May 2003 simultaneously with the FUSE observations using the gratings G140L and G230L.
The spectra cover the wavelength range $1150 - 3200\,\mathrm{\AA}$.
The continuum is estimated considering galactic extinction, Lyman limit absorption, and the intrinsic spectral energy distribution of the QSO.
Shortwards of the Lyman limit break the contribution of the optical depth of the LLS decreases due to the $\nu^{-3}$ dependence of the hydrogen photoionization cross section.
The strength of the break is proportional to the \ion{H}{i} column density of the systems \citep[see][]{mollerjakobsen1990}, which have been measured by \citet{vogelreimers1995}.

The galactic extinction towards HS~1700+6416 is about $E(B-V) = 0.025$ following \citet{schlegeletal1998}, whose values are based on dust maps created from COBE and IRAS infrared data.
X-ray observations of this quasar carried out by \citet{reimersetal1995} yield a hydrogen column density of $N_{\ion{H}{i}} = (2.9 \pm 1.3)\cdot 10^{20}\,\mathrm{cm}^{-2}$.
Following \citet{diplassavage1994}, this corresponds to $E(B-V) = 0.06$.
Though the latter value is derived from direct observations of this line of sight, it is based on the measurement of \ion{H}{i} only assuming a galaxy-wide gas-to-dust ratio, while \citet{schlegeletal1998} use a combination of \ion{H}{i} and the dust distribution.
Therefore, we adopt the value $E(B-V) = 0.025$ from \citet{schlegeletal1998}.

The corrected flux is fitted with a power law $f_{\nu} \propto \nu^{\,\alpha}$ representing the intrinsic QSO continuum.
The best fit succeeds using two power laws with a break at $2000\,\mathrm{\AA}$. 
The best fit yields $\alpha = -1.23$ at $\lambda < 2000\,\mathrm{\AA}$ and $\alpha = -1.54$ at $\lambda > 2000\,\mathrm{\AA}$ using the extinction curve of \citet{cardellietal1989}.
Comparing the STIS spectra to the FOS data analyzed by \citet{vogelreimers1995} taken in December 1991 the flux level is the same in the two datasets down to about $2000\,\mathrm{\AA}$. 
For this wavelength range the spectral index is in agreement with that derived from the older data.
The continuum rises more steeply with decreasing wavelength in the FOS spectrum. At $1200\,\mathrm{\AA}$ the flux level of the STIS data is depressed by a factor of about two in comparison to the FOS data.
Between the two exposures in the course of the observations with FUSE separated about a year, the quasar flux increased about a factor $\sim 3$ in the far-UV.
This means HS~1700+6416 is highly variable in the intrinsic EUV even on relatively short time scales \citep{reimersetal2005b}. 
Thus, in order to extrapolate a reliable continuum it is very important to have simultaneous STIS and FUSE observations.
Fig.\ \ref{continuum} illustrates that the FUSE and STIS portion of the spectrum taken in 2003 are well matched.

A crucial point in the continuum extrapolation is the choice of the extinction curve especially because of its steep rise in the UV.
Adopting the analytic formula of \citet{cardellietal1989} we find a flux level at $1000\,\mathrm{\AA}$ of $\sim 2.69\cdot 10^{-15}\,\mathrm{erg\,s}^{-1}\,\mathrm{cm}^{-2}\mathrm{\AA}^{-1}$.
Tests with the reddening value derived from the X-ray observations \citep{reimersetal1995} lead to a flux at $1000\,\mathrm{\AA}$ depressed by $\sim 25$\,\% (see Fig.\ \ref{continuum}).
Adopting the analytical expression of \citet{pei1992} an the same reddening value leads to a flux level slightly above the original extrapolation.
But since the FUSE spectrum is extremely noisy especially in the lower wavelength range, the dominating uncertainties result from the profile fitting.
The choice of the flux level affects the results only marginally.
The apparent optical depth method is more sensitive to the continuum level.
Adopting the normalization according to the lower continuum level given in Fig.\ \ref{continuum} would lower the $\eta$ values of individual bins by roughly $0.2\,\mathrm{dex}$ near $1000\,\mathrm{\AA}$.
However, we are confident that the continuum extrapolation using the \citet{cardellietal1989} extinction curve and the $E(B-V)$ from \citet{schlegeletal1998} represents a reasonable approximation of the real continuum.

\begin{figure*}
  \centering
  \resizebox{\hsize}{!}{\includegraphics[bb=35 570 545 755,angle=0,clip=]{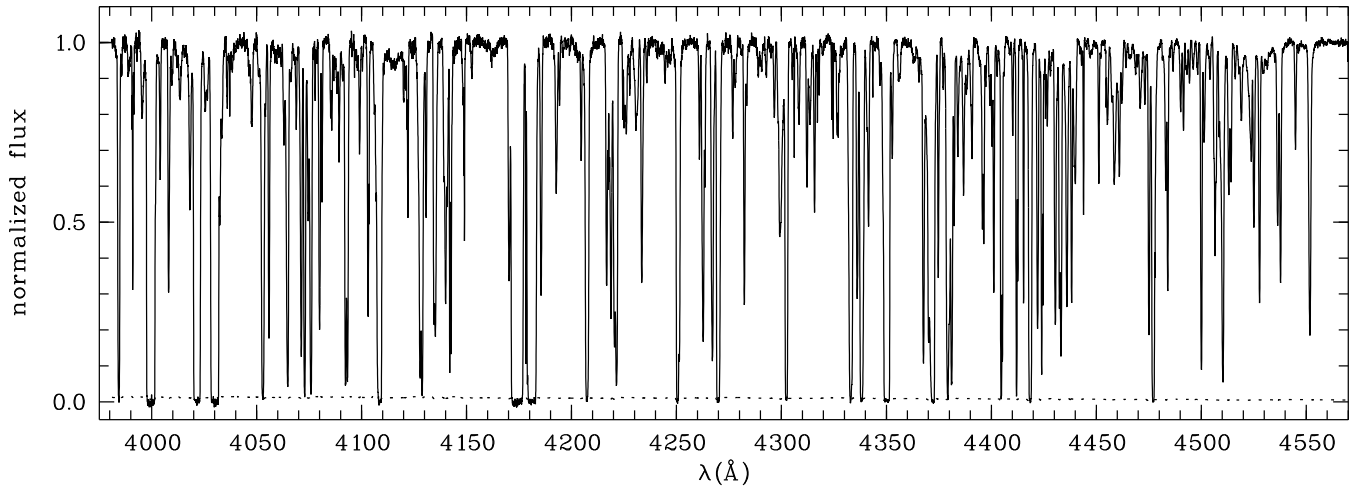}}
  \caption{The Keck data in the range $z > 2.27$. The data normalization is performed simultaneously with the estimation of the line parameters in course of the profile-fitting procedure.
  }
  \label{keck}
\end{figure*}

The continuum in the Keck portion of the spectrum is estimated in the course of the line profile-fitting procedure.
The line-fitting program CANDALF developed by R.\ Baade performs the Doppler profile fit and the continuum normalization simultaneously.
Thus, the continuum determination in the Ly$\alpha$ forest is more reliable in comparison to an a priori continuum definition.
The normalized Keck spectrum in the wavelength range corresponding to the \ion{He}{ii} data is shown in Fig.\ \ref{keck}.

%******************************************************************************
\section{Metal line absorption} \label{mls}

The spectrum of HS~1700+6416 is characterized by a large number of metal line absorption features in the optical spectral range \citep{petitjeanetal1996, trippetal1997, simcoeetal2002, simcoeetal2006} as well as in the UV \citep{reimersetal1992, vogelreimers1993, vogelreimers1995, koehleretal1996}.
Since a significant contribution of absorption by metals is supposed to be present even in the far-UV, the expected metal lines in the FUSE portion have to be taken into account.
Therefore, in the first paper of this series \citep{fechneretal2005a} we construct photoionization models of the metal line systems with the main purpose to derive a prediction of the metal line features expected to arise in the FUSE spectral range.
In the following the main results of \citet{fechneretal2005a} are summarized and we compare the predicted far-UV metal lines to the FUSE data.
For details of the modelling procedure we refer to \citet{fechneretal2005a}.

\begin{figure*}
  \centering
  \resizebox{\hsize}{!}{\includegraphics[bb=25 52 560 777,clip=]{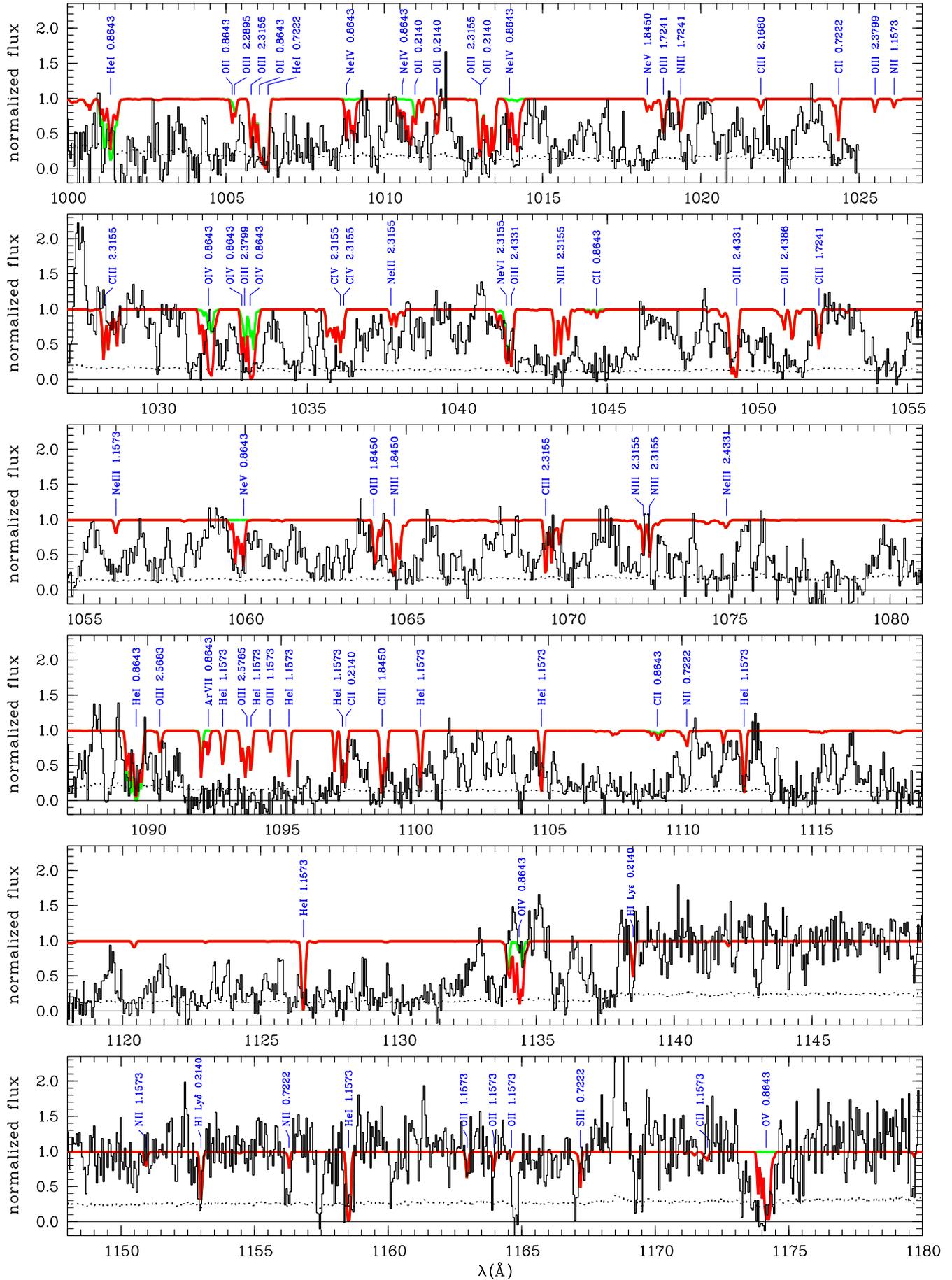}}
  \caption{Predicted metal line spectrum (smooth solid line) in comparison to the data (histogram-like solid line).
The dotted line represents the error of the observational data.
    The predicted features are identified given the ion and the redshift of the system where they arise.
  }
  \label{metals}
\end{figure*}

On the basis of the modelled column densities a metal line spectrum for the wavelength range $1000-1180\,\mathrm{\AA}$ covered by FUSE is predicted. 
These lines \citep[cf.\ Fig.\ 7 of][]{fechneretal2005a} are overlayed to the observed FUSE data in Fig.\ \ref{metals}.
It can be seen, that some rather strong features can be identified as metal lines.
Obviously, the prediction is consistent with the data for most of the lines.

In case of the LLS at $z=0.8643$, however, strong features of oxygen are predicted, but the observed spectrum shows less absorption.
In particular, \ion{O}{iv} is overestimated, as can be seen at $\sim 1032\,\mathrm{\AA}$ and $1134\,\mathrm{\AA}$ in Fig.\ \ref{metals}.
The model of this system is based on the \citet[][HM]{haardtmadau2001} UV background.
However, the spectral energy distribution of a starburst galaxy adopted from \citet[][SB]{bruzualcharlot1993} leads to a reasonable model as well \citep{fechneretal2005a}.
Therefore, another prediction is computed considering the results from the SB model for the $z = 0.8643$ system.
The resulting metal line spectrum is also shown in Fig.\ \ref{metals}.
The main differences in the predicted lines are a decrease of the neon absorption (\ion{Ne}{iii}, \ion{Ne}{iv}) in case of the SB model, but slightly stronger \ion{He}{i} features.
The predicted \ion{O}{iv} features are consistent with the observed data.
However, the strong feature at $1174\,\mathrm{\AA}$ is mainly due to \ion{O}{v} absorption from the $z = 0.8643$ system according to the HM model, while no \ion{O}{v} absorption at all is predicted by the SB model.
We find no alternative identification for this strong feature either proposed by the metal line system models or as part of the Lyman series of a strong, metal-free, low redshift system.
Furthermore, no interstellar absorption line is expected at this wavelength.
We conclude that also the energy distribution of a starburst galaxy does not provide an optimum model of the $z = 0.8643$ system.
Improvement of the photoionization model by adopting a different ionizing background is difficult since only few transitions are observed in the optical.
The vast majority of lines constraining the model are located in the UV, where the data suffers from low $S/N$ \citep[see also][]{fechneretal2005a}.
When we analyze the \ion{He}{ii} Ly$\alpha$ forest taking into account the metal line prediction, we will reconsider the lines of the $z = 0.8643$ systems with different strengths dependent on the adopted model (especially those of \ion{Ne}{iv} and \ion{Ne}{v}) and discuss possible effects on the results.

\begin{table}
  \caption[]{Transitions of galactic H$_2$ that are observed unblended in the FUSE spectrum. The atomic data (rest wavelength $\lambda_{0}$ and oscillator strength $f$) are adopted from \citet{abgralletal1993a, abgralletal1993b}. The column density is estimated assuming $b = 5.0\,\mathrm{km\,s}^{-1}$ and $v = -45.7\,\mathrm{km\,s}^{-1}$.}
  \label{h2}
  $$ 
  \begin{array}{c c c c}
    \hline\hline
    \noalign{\smallskip}
    \mathrm{transition} & \lambda_{0}\, (\mathrm{\AA}) & f & \log N \\
    \noalign{\smallskip}
    \hline
    \noalign{\smallskip}
       \mathrm{L\,(3-0)\,R\,(0)} & \ 1062.882 \ & 0.0178 \ & 16.00 \pm 0.87 \\
       \mathrm{L\,(8-0)\,R\,(1)} & 1002.453 & 0.0183 & 17.83 \pm 0.25 \\
       \mathrm{L\,(3-0)\,R\,(1)} & 1063.460 & 0.0119 & \\
       \mathrm{W\,(0-0)\,R\,(2)} & 1009.023 & 0.0156 & 14.72 \pm 0.30 \\
       \mathrm{W\,(0-0)\,Q\,(2)} & 1010.938 & 0.0245 & \\
       \mathrm{W\,(0-0)\,P\,(2)} & 1012.170 & 0.0055 & \\
       \mathrm{L\,(7-0)\,R\,(2)} & 1014.977 & 0.0190 & \\
       \mathrm{L\,(7-0)\,P\,(2)} & 1016.458 & 0.0102 & \\
       \mathrm{L\,(6-0)\,P\,(2)} & 1028.106 & 0.0105 & \\
       \mathrm{L\,(5-0)\,R\,(2)} & 1038.686 & 0.0165 & \\
       \mathrm{L\,(4-0)\,P\,(2)} & 1053.283 & 0.0090 & \\
       \mathrm{L\,(8-0)\,R\,(3)} & 1006.413 & 0.0158 & 15.49 \pm 0.30 \\
       \mathrm{W\,(0-0)\,R\,(3)} & 1010.129 & 0.0138 & \\
       \mathrm{L\,(6-0)\,R\,(3)} & 1028.983 & 0.0173 & \\
       \mathrm{L\,(5-0)\,R\,(3)} & 1041.157 & 0.0159 & \\
    \noalign{\smallskip}
    \hline
  \end{array}
  $$ 
\end{table}

Absorption from galactic molecular hydrogen is expected to arise additionally in the FUSE spectral range.
Templates for the H$_2$ features with rest wavelengths in the far-UV including Lyman and Werner transitions
and a detailed description of how to identify and eliminate them can be found in \citet{mccandliss2003}.
Since the proposed procedure requires the detection of a sufficient number of undisturbed galactic H$_2$ features, it is inapplicable to our data.
Instead, we proceed as follows:
The templates from \citet{mccandliss2003} are used to easily identify H$_2$ lines, considering that they may be slightly shifted in velocity space. 
Then, we select unblended features in the FUSE spectrum and fit them with Doppler profiles to estimate the line parameters.
The selected transitions and the fitted line parameters are given in Table \ref{h2}. 
All considered rotational transitions (ground state to $J'' = 0, 1, 2, 3$) can be modelled with a Doppler parameter $b = 5.0$ at a velocity $v = -45.7\,\mathrm{km\,s}^{-1}$.
The derived column densities for every $J''$ are also listed in Table \ref{h2}.
Because of the few observed unblended lines and the poor data quality, the column densities, especially those of the lower rotational transitions, are very uncertain.
Nevertheless, the derived values are roughly consistent with the column densities derived by \citet{richteretal2001, richteretal2003} for high- and intermediate-velocity clouds, respectively, towards extragalactic sources.
In the final step, all transitions within the observed spectral range with $J'' = 0, 1, 2, 3$ are included into our model.

In the following we treat the H$_2$ model like the predicted metal absorption lines.
Unless explicitly noted we refer to the predicted metal line absorption as well as galactic H$_2$ when talking about metal lines.
Using either the profile-fitting procedure or the apparent optical depth method all additional lines are considered in the analysis of the \ion{He}{ii} Ly$\alpha$ forest.
Besides the metal lines also the interstellar lines \ion{C}{ii} $\lambda 1036$ and \ion{Ar}{i} $\lambda\lambda 1048, 1067$ are included.

\begin{figure*}
  \centering
  \resizebox{\hsize}{!}{\includegraphics[bb=35 385 520 755,clip=]{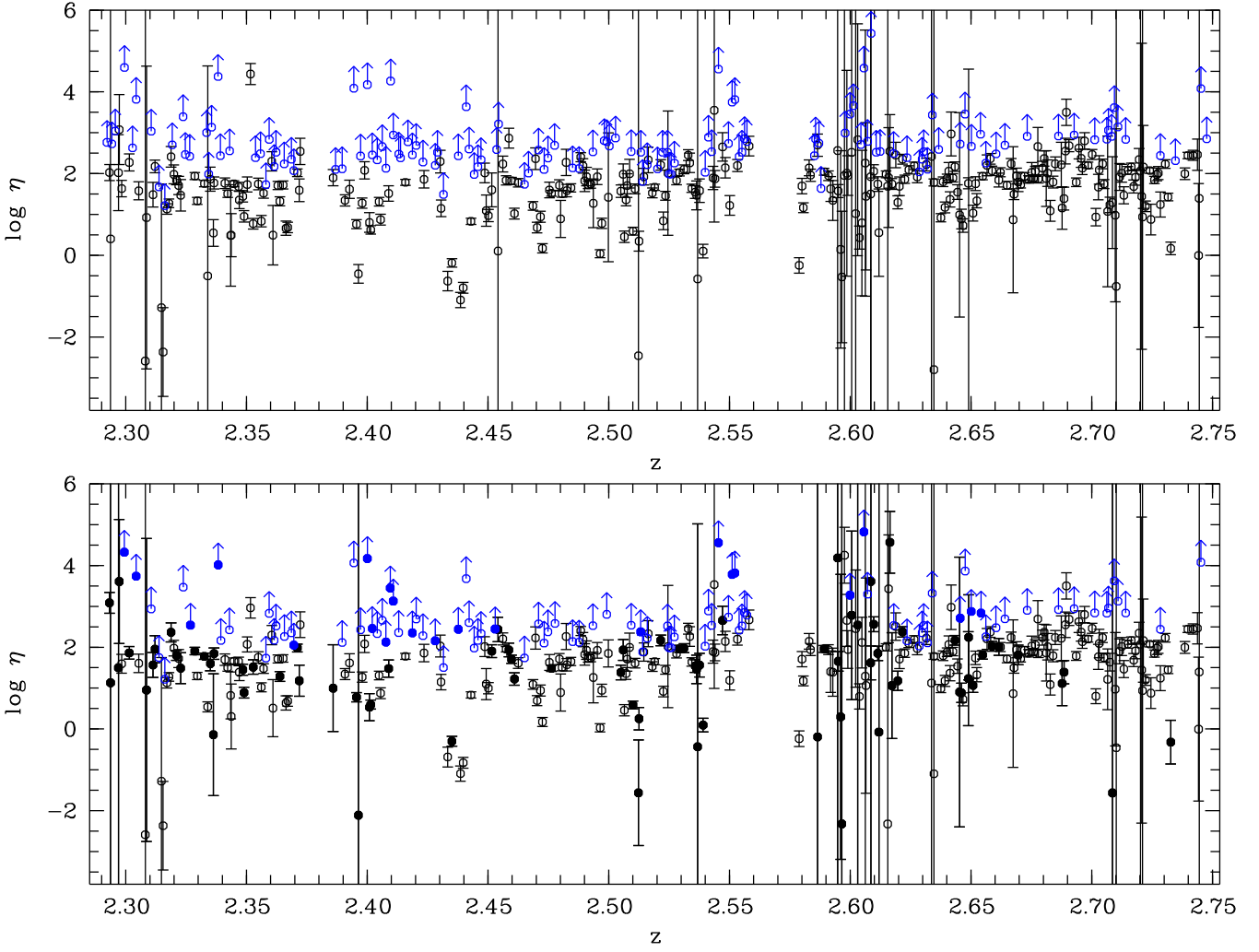}}
  \caption{Comparison of the $\eta (z)$ distribution ignoring metal line and galactic H$_2$ absorption (upper panel) and taking them into account (lower panel). Absorbers affected by extra absorption are marked as filled circles.
Lower limits are based on \ion{He}{ii} features without detected \ion{H}{i} counterparts. Their Doppler parameters are chosen to be $b = 27.0\,\mathrm{km\,s}^{-1}$.
  }
  \label{simple_prognose}
\end{figure*}

We roughly estimate how strongly the FUSE spectral range might be affected by lines of the higher Lyman series lines of the \ion{H}{i} Ly$\alpha$ forest at $z \lesssim 0.2$.
The model spectrum derived in \citet{fechneretal2005a} is compared to the STIS UV data covering the wavelength range $1230 - 1550\,\mathrm{\AA}$ used to constrain the models.
Assuming all unidentified, strong features are low redshift Ly$\alpha$ lines, the expected positions of the higher order Lyman series lines are computed.
Most of the systems are not expected to contribute to the absorption in the FUSE spectral range since higher order Lyman series lines located redwards of the \ion{He}{ii} forest are detected only very weakly.
However, three systems at $z = 0.1240$, $0.0809$, and $0.0237$ possibly affect the \ion{He}{ii} Ly$\alpha$ forest.
The Ly$\beta$ feature of the system at $z = 0.1240$ is blended with Ly$\delta$ of the metal line system at $z = 0.2140$ both located at $\sim 1153\,\mathrm{\AA}$.
The corresponding Ly$\gamma$ and $\delta$ features are expected at $1093.1\,\mathrm{\AA}$ and $1067.5\,\mathrm{\AA}$, respectively. 
From the system at $z = 0.0809$ absorption is expected at $1108.7\,\mathrm{\AA}$ (Ly$\beta$) and $1051.2\,\mathrm{\AA}$ (Ly$\gamma$), while the Ly$\beta$ features of the $z = 0.0237$ systems should be located at $1050.0\,\mathrm{\AA}$.
As an be seen from Fig.\ \ref{simple_prognose}, no remarkably high $\eta$ values are detected at $z \sim 2.456$, $2.460$, $2.514$, and $2.650$, corresponding to the wavelength positions discussed above.
Therefore, biases in $\eta$ due to higher order Lyman series lines of the low redshift \ion{H}{i} forest should be negligible.

The importance of metal line absorption in the FUSE spectral range is illustrated in Fig.\ \ref{simple_prognose}.
The upper panel shows the redshift distribution of $\eta$ estimated by fitting Doppler profiles to the \ion{H}{i} and \ion{He}{ii} Ly$\alpha$ forest ignoring metal line absorption (for details of the analysis procedure see Sect.\ \ref{methods}), while metal line features have been considered in the lower panel.
Lower limits indicate \ion{He}{ii} absorbers without detected \ion{H}{i} counterpart.
About 32\,\% of the apparent \ion{He}{ii} features have no \ion{H}{i} counterpart when metal line absorption is ignored.
The consideration of metal lines reduces the number of lower limits significantly, since several absorption features can be identified as metal lines.
In this case only 25\,\% additional \ion{He}{ii} lines are needed.
Data points that have been biased by metal line absorption, are marked in the lower panel of Fig.\ \ref{simple_prognose}.
In total roughly 27\,\% of the measured values are affected.
For 19\,\% of the data points, $\eta$ changes by more than $\Delta\log\eta = 0.01$ when extra absorption is taken into account.
Considering only features of galactic H$_2$, about 6\,\% of the data points would have been biased, i.e.\ the major contribution comes from the predicted metal line absorption.
There are no indications of unusually high column density ratios at the position of metal lines arising from the $z = 0.8643$ system.

The median column density ratio decreases from $\log\eta = 1.99$ if metal lines are neglected to $1.93$ if they are taken into account.
Consequently, the consideration of metal line absorption has no severe influence on the general statistical properties of the $\eta$ distribution.
The column density ratio of a single absorber, however, can be distorted up to an order of magnitude.

%******************************************************************************
\section{Simulations}\label{simulations}

Two different methods have been used so far to analyze the data of the \ion{He}{ii} Ly$\alpha$ forest towards HE~2347-4342.
\citet{krissetal2001} and \citet{zhengetal2004} fit line profiles to the observed absorption features.
The use of profile fits assumes the absorption features to arise from discrete clouds in the IGM, which is certainly an oversimplification.
The alternative method measuring the apparent optical depth was also applied to the HE~2347-4342 data \citep{shulletal2004}.
In this case the column density ratio $\eta$ is estimated by the ratio of the \ion{He}{ii} and \ion{H}{i} optical depth per redshift or velocity bin.
Since $\eta$ is defined per bin, the assumption of discrete absorbers can be dropped in favour of a continuous medium, where the features are due to density fluctuations.  
A third approach fitting the optical \ion{H}{i} Ly$\alpha$ forest directly to the UV spectrum in order to derive $\eta$ has been introduced by \citet{fechnerreimers2005}.

In order to study possible effects of the analysis method on the results we create artificial datasets.
By evaluating the artificial spectra we investigate how the results are influenced by the applied method.
For both methods we recognize problems which prevent the accurate reconstruction of the presumed $\eta$ value.
In the following we briefly summarize the methodical procedures, then describe the generation of the artificial data, and discuss the implications of the profile-fitting procedure and the difficulties using the apparent optical depth method.

%------------------------------------------------------------------------------
\subsection{Methods}\label{methods}

In preparation of the profile fit of the \ion{He}{ii} spectrum, we first identify all \ion{H}{i} Ly$\alpha$ features in the optical data and estimate their parameters by fitting Doppler profiles.
In the case of strongly saturated Ly$\alpha$ lines higher orders of the Lyman series up to Ly$\epsilon$ are considered as well to derive the parameters.
Then, the \ion{He}{ii} Ly$\alpha$ forest is fitted fixing the redshift and the Doppler parameter derived from the corresponding hydrogen lines.
The latter constraint implies, that we assume pure turbulent broadening.
\citet{zhengetal2004} made the attempt to derive the dominant broadening mechanism from the HE~2347-4342 data fitting unblended \ion{He}{ii} features with a free $b$-parameter. 
They found a velocity ratio $\xi = b_{\mathrm{He}}/b_{\mathrm{H}}$ close to $1.0$, suggesting turbulent broadening to be dominant.
However, \citet{fechnerreimers2005} demonstrate that thermal broadening is important for part of the absorbers and can lead systematically to lower $\eta$ values for high density \ion{H}{i} absorbers.

The procedure of the apparent optical depth method applied to \ion{He}{ii} Ly$\alpha$ forest data as described by \citet{shulletal2004} will be summarized briefly:
The column density ratio is replaced by $\eta = 4\cdot\tau_{\ion{He}{ii}}/\tau_{\mathrm{\ion{H}{i}}}$ \citep[][more exactly the factor 4 represents the ratio of the rest wavelengths $\lambda_{0,\,\mathrm{\ion{H}{i}}}/\lambda_{0,\,\ion{He}{ii}} = 1215.6701/303.7822$]{miraldaescude1993}, where the optical depths $\tau = -\ln F$ are measured per bin.
In order to obtain physically reasonable values, bins with normalized fluxes $F$ above unity are masked out as well as bins with flux values below zero.
Additionally, bins with flux values within $1\,\sigma$ from unity or zero are not considered, since they cannot be distinguished statistically from the continuum or zero, respectively.

%------------------------------------------------------------------------------
\subsection{Creating artificial datasets}

Based on the statistical properties of the \ion{H}{i} Ly$\alpha$ forest as observed towards HS~1700+6416, a sample of lines is generated.
The column density distribution function with $\beta = 1.5$ is adopted from \citet{kirkmantytler1997}. Values in the range $11.0 \le \log N \le 18.0$ are simulated.  
Our observed line sample yields $\beta = 1.51 \pm 0.05$ for absorbers with $12.8 \le \log N \le 16.0$.
The Doppler parameter distribution is described by a truncated Gaussian 
\[
       \frac{\mathrm{d}\mathcal{N}(b)}{\mathrm{d}b} = 
       \left\{ \begin{array}{ll}
	 \displaystyle A \cdot \exp\left( -\frac{(b - b_{0})^{2}}{2\,\sigma_{b}^{2}}\right) & \quad b \ge b_{\mathrm{min}}\\ 
	 \displaystyle  0 & \quad  b < b_{\mathrm{min}}
       \end{array} \right.
\]
with $b_{0} = 27\,\mathrm{km\,s}^{-1}$, $\sigma_{b} = 8.75\,\mathrm{km\,s}^{-1}$, and $b_{\mathrm{min}} = 10\,\mathrm{km\,s}^{-1}$ as observed towards HS~1700+6416 in good agreement with \citet{huetal1995}.
In addition, the parameters of the simulated line sample are correlated by $b_{\mathrm{min}} = (1.3\cdot(\log N - 12.5) + 10.5)\,\mathrm{km\,s}^{-1}$ following \citet{misawa2002}.
The artificial spectrum covers the redshift range $2.292 < z < 2.555$, which corresponds to wavelengths shorter than $1080\,\mathrm{\AA}$ in the FUSE spectrum.
No number density evolution is considered along this redshift interval.
The minimal distance between two lines is $\mathrm{d}z = 0.0001$ as observed towards HS~1700+6416.
In the given redshift range the separation corresponds to $\sim 20\,\mathrm{kpc}$.
The resolution of $R = 40\,000$ and the signal-to-noise ratio of $S/N \sim 100$ are chosen to match the characteristics of high-resolution spectra taken with VLT/UVES or Keck/HIRES.

The \ion{He}{ii} Ly$\alpha$ forest is computed based on the artificial \ion{H}{i} lines using $\eta = 80$, the mean value found by \citet{krissetal2001}.
Considering a temperature of $10^{4}\,\mathrm{K}$ the \ion{He}{ii} Doppler parameter is computed as
\begin{equation}
 b_{\ion{He}{ii}} = \sqrt{\,b_{\mathrm{\ion{H}{i}}}^{2} - 2 k T \left( \frac{1}{m_{\mathrm{H}}} - \frac{1}{m_{\mathrm{He}}} \right)}\,\mbox{.}
\end{equation}
The assumed temperature is cooler than the average $2\cdot 10^{4}\,\mathrm{K}$ measured by \citet{ricottietal2000}, but consistent with $b_{\mathrm{min,\,\ion{H}{i}}} = 10\,\mathrm{km\,s}^{-1}$. 
Test calculations with higher temperatures indicate no difference in the results. 
Since the IGM is expected to have a distinctive temperature distribution, a constant temperature model is certainly an oversimplification, but serves to get an impression of the possible effects assuming pure turbulent broadening.
For comparison we compute as well a \ion{He}{ii} dataset with $b_{\ion{He}{ii}} = b_{\mathrm{\ion{H}{i}}}$.
The resolution of $R = 15\,000$ and the signal-to-noise ratio of $S/N \sim 5$ are chosen to match the typical values of real FUSE data.

Since the artificial spectra are generated from statistical assumptions alone, they provide only a simple approach to investigate the applicability of the methods.
More realistical spectra based on hydrodynamical simulations are presented by \citet{boltonetal2006}.
They find that the column density ratio can be estimated confidently by a line profile-fitting procedure \citep[but see discussion below and in][]{fechnerreimers2005}.
Furthermore, their simulations permit the investigation of different UV backgrounds, whereas the simple statistical approach introduced here can provide only limited information about basic problems of the analysis procedures.

%------------------------------------------------------------------------------
\subsection{Profile fitting analysis}

\begin{figure}
  \centering
  \resizebox{\hsize}{!}{\includegraphics[bb=28 415 460 740,clip=]{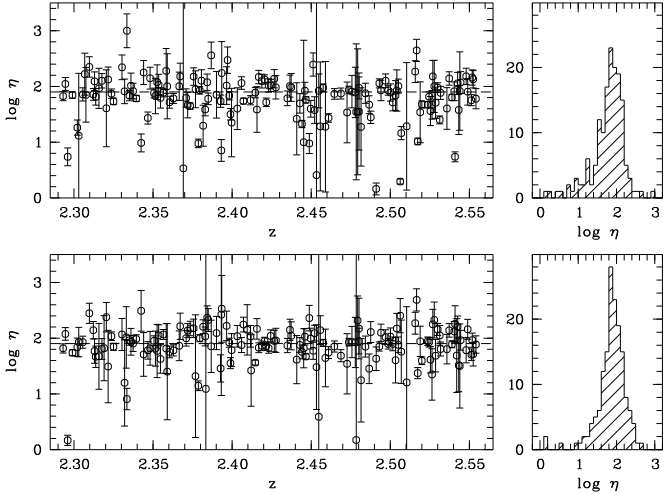}}
  \caption{Distributions of the column density ratio $\eta$ derived from the artificial data with a Doppler parameter consisting of a thermal and a turbulent part (upper panel) or a pure turbulent $b$-parameter (lower panel) applying a profile-fitting procedure. The dashed line indicates the presumed value $\eta = 80$. In the course of the analysis pure turbulent broadening is assumed in both cases.
  }
  \label{eta_therm_turb}
\end{figure}

Having analyzed the artificial \ion{H}{i} Ly$\alpha$ forest by fitting Doppler profiles, the statistical properties of the sample are recovered.
Under the simulated conditions our sample is complete down to $\log N \sim 12.0$. 
Due to blending effects, the deduced Doppler parameter distribution is broadened by $28\,\%$.
The \ion{He}{ii} Ly$\alpha$ forest lines are treated the same way, using the derived \ion{H}{i} parameters, with fixed line redshifts and $b$-parameters.
Features with Doppler parameters $b < 10\,\mathrm{km\,s}^{-1}$ are not considered. 
The resulting $\eta$ values are shown in Fig.\ \ref{eta_therm_turb}.
Both the combined thermal and turbulent as well as the pure turbulent broadened sample show a scatter in $\eta$ of about $0.5\,\mathrm{dex}$. 
Generally, the line sample that contains thermal broadening has more low $\eta$ values.
We find a statistical mean of $\log\eta_{\mathrm{therm}} = 1.77 \pm 0.43$ in comparison to $\log\eta_{\mathrm{turb}} = 1.86 \pm 0.37$ in the pure turbulent case, both still consistent with the presumed value of $1.90309$.
The median, which gives less weight to outliers, is $\log\eta_{\mathrm{therm}} = 1.86$ and $\log\eta_{\mathrm{turb}} = 1.90$, respectively.

There are various reasons for outliers lying outside the range $\log\eta = 1.903 \pm 0.500$. 
Both models show only about $3\,\%$ high values caused by line saturation and blending effects, which can, in principle, produce low as well high $\eta$ values.
In the sample with pure turbulent broadening we find $8\,\%$ low $\eta$ values while the thermal and turbulent broadened sample shows $14\,\%$ low values.
Besides saturation and blending, the uncertainties of the parameters of weak \ion{H}{i} lines resulting in large error bars are a primary reason.
In the case of combined thermal and turbulent broadening some \ion{He}{ii} lines are narrower than the adopted pure turbulent $b$-parameter of \ion{H}{i}.
This effect produces about $25\,\%$ of the low values.
Therefore, if the widths of intergalactic absorbers are not completely dominated by turbulent mechanisms, low $\eta$ values can be caused by the assumption of pure turbulent broadening \citep[see also][]{fechnerreimers2005}.

\begin{figure}
  \centering 
  \resizebox{\hsize}{!}{\includegraphics[bb=42 575 230 770,clip=]{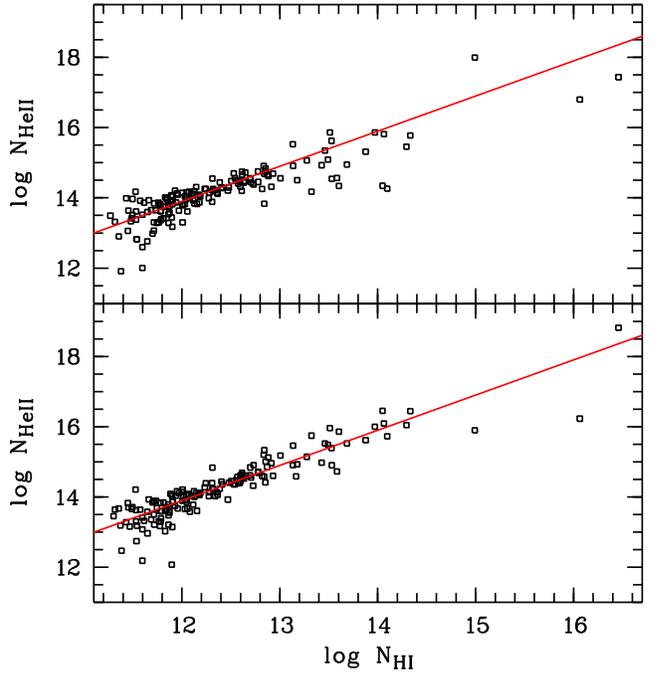}}
  \caption{Recovered \ion{He}{ii} column densities versus \ion{H}{i} column densities for combined thermal and turbulent broadening (upper panel) and in the pure turbulent case (lower panel). The solid line represents the underlying correlation $\eta = 80$. For clarity, measurement errors are not plotted.
  }
  \label{logN_logN}
\end{figure}

In order to find a strategy for avoiding artificial scatter in the $\eta$ distribution, we directly examine the correlation between \ion{H}{i} and \ion{He}{ii} column density as presented in Fig.\ \ref{logN_logN}.
The correlation fits well except for very low column densities ($\log N_{\ion{H}{i}} < 12.0$) where noise makes it difficult to derived accurate line parameters, and also for high column densities, where saturation of hydrogen lines plays a role (features are saturated when $\log N_{\ion{H}{i}} \gtrsim 14.5$ for $b \approx 27\,\mathrm{km\,s}^{-1}$). 
Since the redshift and Doppler parameter of the helium lines are fixed during the fitting procedure, the column densities are derived correctly if the underlying assumptions are correct.
In the case of combined thermal and turbulent broadening the \ion{He}{ii} column densities are significantly underestimated for $\log N_{\ion{H}{i}} \gtrsim 13.0$, caused by the assumption of pure turbulent broadening.
\citet{fechnerreimers2005} argued that the incorrect assumption of pure turbulent broadening affects only lines above this \ion{H}{i} column density.
Thus, a stringent criterion to define a reasonable subsample would be $12.0 \le \log N_{\ion{H}{i}} \le 13.0$.
Assuming pure turbulent broadening as a good approximation, $12.0 \le \log N_{\ion{H}{i}} \le 14.5$ provides a less stringent constraint.

\citet{boltonetal2006} find a distribution similar to that shown in the lower panel of Fig.\ \ref{logN_logN}.
Their lines samples contain only absorber with $\log N_{\ion{H}{i}} > 12.0$ and show enhanced scatter for $\log N_{\ion{H}{i}} > 13.0$ even in the case of a uniform UV background.
Since they assumed turbulent broadening as well, the scatter most likely indicates the effect of thermal line widths.
However, \citet{boltonetal2006} neither discuss the origin of the scatter nor the possible implications of the assumption of pure turbulent broadened lines.

Applying the criteria derived above to the simulated datasets we find the statistical means $\log\eta_{\mathrm{therm}} = 1.85 \pm 0.21$ and $\log\eta_{\mathrm{turb}} = 1.92 \pm 0.21$ for the more stringent constraint ($12.0 \le \log N_{\ion{H}{i}} \le 13.0$).
In the pure turbulent case, the underlying $\eta$ value is recovered well, while it is still underestimated when thermal broadening plays a role.
For both samples the statistical $1\,\sigma$ error has been reduced by a factor $\sim 2$ in comparison to the total sample.
Considering the sample $12.0 \le \log N_{\ion{H}{i}} \le 14.5$ yields $\log\eta_{\mathrm{therm}} = 1.74 \pm 0.39$ and $\log\eta_{\mathrm{turb}} = 1.92 \pm 0.25$, respectively.
Here, the underestimation of $\eta$ for the data including thermal broadening gets worse and the scatter increases, while the mean value and $1\,\sigma$ error for the pure turbulent data do not change significantly.

%------------------------------------------------------------------------------
\subsection{Apparent optical depth method}\label{appopt}

\begin{figure}
  \centering
  \resizebox{\hsize}{!}{\includegraphics[bb=28 440 460 748,clip=]{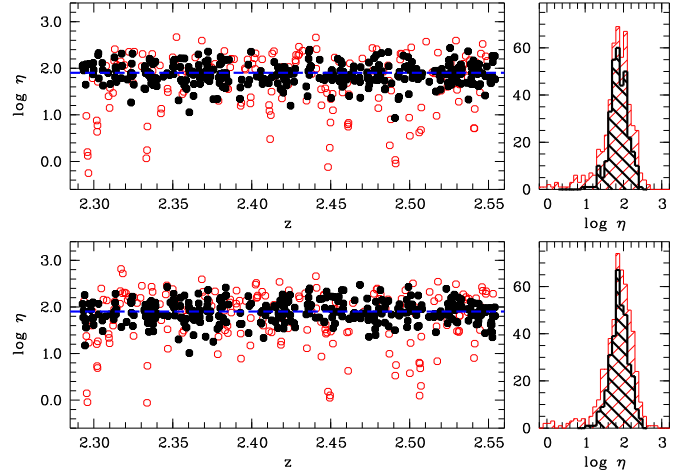}}
  \caption{Distributions of the column density ratio $\eta$ derived from the artificial data with a Doppler parameter consisting of a thermal and a turbulent part (upper panel) or a pure turbulent $b$-parameter (lower panel) using the apparent optical depth method. 
The dashed line indicates the presumed value $\eta = 80$. 
The bin size is $20\,\mathrm{km\,s}^{-1}$, corresponding to $\sim 0.05\,\mathrm{\AA}$ in \ion{He}{ii}. 
Bins with $-2.0 \le \log\tau_{\ion{H}{i}} \le -1.0$ (see text) are denoted as filled circles, all others as open circles. 
The corresponding distributions are indicated as histograms in the right panels.
  }
  \label{eta_therm_turb_bin}
\end{figure}

The same artificial datasets are analyzed applying the apparent optical depth method.
The resulting distribution for a bin size of $\Delta v = 20\,\mathrm{km\,s}^{-1}$, corresponding to $0.05\,\mathrm{\AA}$\, in \ion{He}{ii} wavelengths, is shown in Fig.\ \ref{eta_therm_turb_bin}.
This method leads to a scatter in $\eta$ of about $0.5\,\mathrm{dex}$ for both models as well.
The high values are statistically negligible, since they are only about $3\,\%$ for each sample.
Again we find slightly more low $\eta$ values when thermal broadening is present (roughly $13\,\%$ in comparison to $10\,\%$ in the case of pure turbulent broadening).
This is plausible, since the \ion{He}{ii} optical depth in the wings of a thermally broadened, and therefore narrower feature is systematically lower than in the case of pure turbulence.
In both models the underlying $\eta$ is underestimated.
We find the mean values $\log\eta_{\mathrm{therm}} = 1.80 \pm 0.44$ and $\log\eta_{\mathrm{turb}} = 1.84 \pm 0.41$.
With increased bin size the mean values get even lower ($\log\eta_{\mathrm{therm}} = 1.75 \pm 0.39$ and $\log\eta_{\mathrm{turb}} = 1.77 \pm 0.38$ for $\Delta v = 60\,\mathrm{km\,s}^{-1}$, corresponding to a bin size of $0.20\,\mathrm{\AA}$\, in \ion{He}{ii}). 
Also the median underestimates the real $\eta$ in the combined broadening case because of the excess of low values, even though this effect is smaller.
We find $\log\eta_{\mathrm{therm}} = 1.88$ and $\log\eta_{\mathrm{turb}} = 1.90$ (bin size $20\,\mathrm{km\,s}^{-1}$), in comparison to $\log\eta_{\mathrm{therm}} = 1.81$ and $\log\eta_{\mathrm{turb}} = 1.83$ with $60\,\mathrm{km\,s}^{-1}$ bin size.

\begin{figure}
  \centering
  \resizebox{\hsize}{!}{\includegraphics[bb=35 405 240 770,clip=]{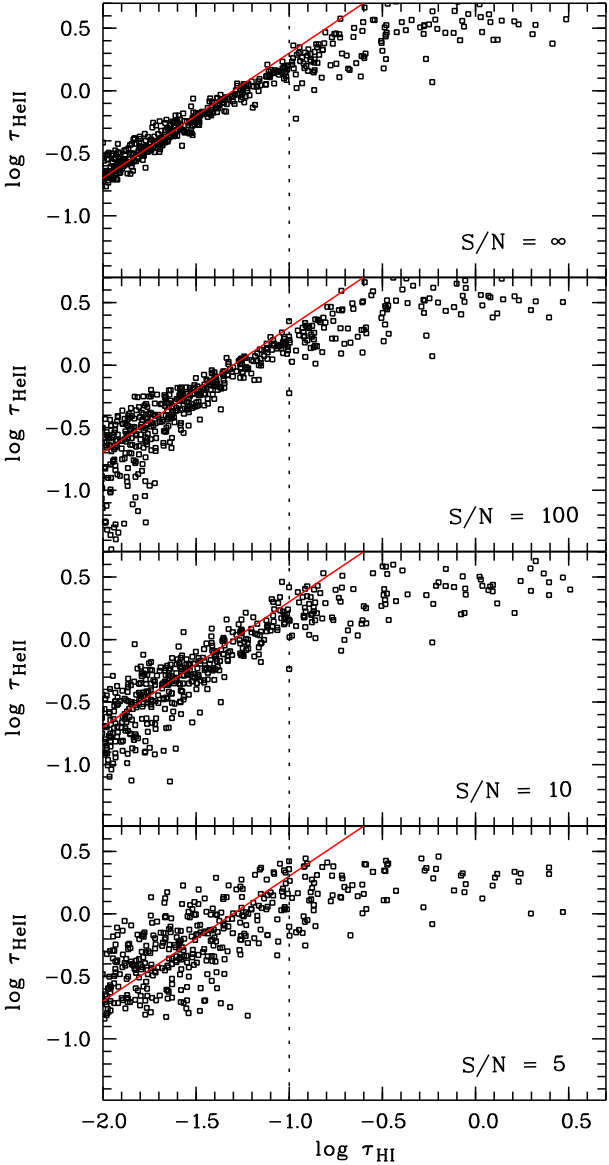}}
  \caption{The \ion{He}{ii} optical depth versus \ion{H}{i} optical depth for simulated data with pure turbulent broadening. 
The specific signal-to-noise of the artificial helium dataset is given in each panel. 
The signal-to-noise ratio of the hydrogen spectrum is $S/N = 100$. 
The upper panel is based on artificial datasets with no noise at all. 
The resolution of the simulated spectra are $R_{\ion{H}{i}} = 40\,000$ and $R_{\ion{He}{ii}} = 15\,000$, respectively. 
Each point represents a bin of $20\,\mathrm{km\,s}^{-1}$. 
The solid lines indicate $\eta = 80$. 
  }
  \label{tau_SN}
\end{figure}

The correlation between low \ion{H}{i} optical depth, referred to as voids, and high $\eta$ values as found by \citet{shulletal2004} can also be seen in the simulated data even though it is not present in the underlying spectra.
For further investigation of this point it is instructive to regard $\eta$ as the correlation between $\tau_{\ion{H}{i}}$ and $\tau_{\ion{He}{ii}}$.
This is shown in Fig.\ \ref{tau_SN} in the pure turbulent case for different signal-to-noise ratios of the \ion{He}{ii} data (100, 10, and 5, while $S/N = 100$ is constant for the \ion{H}{i} spectra) and an example without any noise.
 
At the given $S/N$ of the \ion{H}{i} spectrum ($=100$), optical depths down to $\log\tau_{\ion{H}{i}} > -2.0$ can be measured reliably, at least in principle. 
If only bins detached by more than $2\,\sigma$ from unity are considered, the limit is $\log\tau_{\ion{H}{i}} > -1.7$. 
The scatter at low optical depths (e.g.\ for $\log\tau_{\ion{H}{i}} \lesssim -1.7$ in the case of $S/N = 100$ for \ion{H}{i} as well as \ion{He}{ii}) is to due the noise.
In order to avoid a contamination by noise-affected bins, the constraints defining the subsample should be chosen carefully.
As can be seen from Fig.\ \ref{tau_SN}, the detection limit of the \ion{He}{ii} optical depth depends on the signal-to-noise, of course.
In case of $S/N = 5$ the restriction is $\log\tau_{\ion{He}{ii}} > -0.65$, or $\log\tau_{\ion{He}{ii}} > -0.3$ for the $2\,\sigma$ constraint.

Even the analysis of the spectra without noise leads to scatter around the underlying relation.
Furthermore, bins with an optical depth above $\log\tau_{\ion{H}{i}} > -1.0$ (indicated by the vertical, dotted line in Fig.\ \ref{tau_SN}) yield \ion{He}{ii} optical depths clearly beneath the expected values.
This flattening is due to line saturation, since saturated absorption features become broader and more smeared out at finite resolution.
The difference in the profile does not only affect the saturated core but as well the wings of the line.
Consequently, the flattening starts at relatively low hydrogen optical depths. 
Thus, a direct comparison of the apparent optical depths above a certain limit is not possible. 

\citet{savagesembach1991} give a detailed description of the method and applicability using the apparent optical depth. 
They emphasize the possibility of finding line saturation in the analysis of line doublets.
Hidden saturation can be seen if the apparent column density $N_{a}(v) \propto \tau_{a}(v)/f\,\lambda_{0}$ of a doublet pair differs with respect to the considered component.
Since line saturation is important in the whole FUSE spectrum, the apparent optical depth method is not applicable to bins with $\tau_{\ion{H}{i}} > 0.1$ if the underlying $\eta$ is roughly 80.
Adopting a $1\,\sigma$ detection limit of $\tau_{\ion{H}{i}} > 0.01$ there is $1\,\mathrm{dex}$ of hydrogen optical depth left, allowing the apparent optical depth to be applied reasonably.

\citet{foxetal2005} investigated the effect of noise on the apparent column density.
They found that the apparent optical depth method will likely overestimate the true column density when applied to data with low $S/N$.
Since the FUSE spectra are very noisy ($S/N \approx 5$), the results obtained by an apparent optical depth method have to be considered carefully.

The bins selected by $-2.0 \le \log\tau_{\ion{H}{i}} \le -1.0$ are marked in Fig.\ \ref{eta_therm_turb_bin} and the overlay in the right panels shows their $\eta$ distribution.
Both the samples (pure turbulent and combined line widths) contain virtually no low $\eta$ values and the fraction of high values is below 5\,\% in each case.
As expected, the mean values approach the underlying $\eta$ and the scatter decreases.
We find $\log\eta_{\mathrm{therm}} = 1.88 \pm 0.24$ and $\log\eta_{\mathrm{turb}} = 1.90 \pm 0.24$, respectively, with a bin size of $20\,\mathrm{km\,s}^{-1}$.
Fitting a Gaussian $A\cdot \exp(-(\log\eta - \log\eta_{0})^{2}/2\,\sigma^{2})$ to the $\log\eta$ distribution of the subsample leads to $\log\eta_{0} = 1.90 \pm 0.01$ and $\sigma = 0.237 \pm 0.009$ in the case of pure turbulence. 
In the case of combined thermal and turbulent broadening we find $\log\eta_{0} = 1.89 \pm 0.16$ and $\sigma = 0.213 \pm 0.012$.
In comparison to the mean value of the sample without any noise, which is $1.89 \pm 0.07$ in the pure turbulent case, these values demonstrate that the scatter in the $\eta$ distribution is mainly due to the noise level of the data.
With the present FUSE data ($S/N \sim 5$) an $\eta$ distribution with a scatter of roughly 13\,\% can be expected following the simulations.

%------------------------------------------------------------------------------
\subsection{Comparing profile fitting and apparent optical depth}

The investigation of artificial spectra by fitting Doppler profiles or applying the apparent optical depth method exhibits differences of the robustness and applicability of the diagnostics.
Concerning the apparent optical depth, line saturation in combination with the detection limit given by the noise leads to a relatively small interval of 1\,dex in the hydrogen optical depth ($0.01 \le \tau_{\ion{H}{i}} \le 0.1$), where the method can be applied reliably.
Besides limited resolution noise is the most crucial limitation, since measuring the apparent optical depth per velocity bin is directly sensitive to the noise level, while the impact of the assumption of pure turbulent line width is less significant (the deviation is roughly 1\,\%).

Since position and line width are fixed, noise has less influence using the standard profile-fitting method.
Therefore, even saturation is less problematic.
Hydrogen starts to be saturated at column densities $\log N_{\ion{H}{i}} \gtrsim 14.5$ and normally higher order Lyman series lines can be used to estimate the \ion{H}{i} line parameters accurately in case of strong saturation.
Thus, even saturated \ion{He}{ii} features may be evaluated reasonably.
However, in the crowded Ly$\alpha$ forest the number of components may be ambiguous leading to larger uncertainties.
Another point is, in analyzing the simulated data, we know that the correct profile function is applied since the datasets are created using it.
When dealing with real data, the observed features can be described only approximately.

A sample restricted to lines that can be clearly identified and are insensitive to effects of thermal broadening is expected to yield the most reliable results.
The selected lines have to be strong enough to be fitted unambiguously but are below the sensitivity limit for thermal line widths ($12.0 \le \log N_{\ion{H}{i}} \le 14.5$ or even less), since a line sample under the simplistic assumption of pure turbulent broadening will underestimate the true $\eta$ values.
According to \citet{fechnerreimers2005} features with $\log N_{\ion{H}{i}} \lesssim 13.0$ remain unaffected by effects of thermal broadening.
Therefore a more stringent selection would be $12.0 \le \log N_{\ion{H}{i}} \le 13.0$.

\begin{table}
  \caption[]{The $\eta$ values recovered from the simulated data using the profile fitting or apparent optical depth method, respectively. 
The input value is 80, i.e.\ $\log\eta = 1.90309$. 
The first columns show the values for the profile-fitting analysis constraining the sample to $12.0 \le \log N_{\ion{H}{i}} \le 13.0$ (selected) and without any restriction (unselected).
The columns at the right hand side give the results obtained with the apparent optical depth method (binning $20\,\mathrm{km\,s}^{-1}$) with $-2.0 \le \log\tau_{\ion{H}{i}} \le -1.0$ (selected) and for the total sample (unselected).}
  \label{compilation}
  $$ 
  \begin{array}{l l c c c c}
    \hline\hline
    \noalign{\smallskip}
             &  & \multicolumn{2}{c}{\mathrm{profile~fitting}} & \multicolumn{2}{c}{\mathrm{apparent~optical~depth}}\\
             &  & \langle\log\eta\rangle & \mathrm{median} & \langle\log\eta\rangle & \mathrm{median} \\
            \noalign{\smallskip}
            \hline
            \noalign{\smallskip}
\mathrm{turbulent} & \mathrm{selected}  & 1.92 \pm 0.21 & 1.92 & 1.90 \pm 0.24 & 1.89 \\
                   & \mathrm{unselected}& 1.86 \pm 0.37 & 1.90 & 1.84 \pm 0.41 & 1.90 \\
\mathrm{thermal}   & \mathrm{selected}  & 1.85 \pm 0.21 & 1.87 & 1.88 \pm 0.24 & 1.88 \\
                   & \mathrm{unselected}& 1.77 \pm 0.43 & 1.89 & 1.80 \pm 0.44 & 1.88 \\
            \noalign{\smallskip}
            \hline
         \end{array}
     $$ 
   \end{table}

Table \ref{compilation} lists the recovered $\eta$ values and their statistical $1\,\sigma$ errors using either the profile fitting or the apparent optical depth method.
The median reproduces the underlying value of $\log\eta = 1.903$ within an accuracy of more than 93\,\% in all cases.  
The statistical mean underestimates the column density ratio fitting line profiles, particularly if thermal broadening is present.
This is consistent with the results of \citet{boltonetal2006}, who come to the conclusion that the median gets closer to the underlying value, although they do not discuss possible reasons for this finding.
Furthermore, the scatter is reduced by more than 40\,\% if only the selected subsamples are considered.

Following these results, we expect to find a scatter in $\log\eta$ of roughly 20\,\% by analyzing the observed spectra without any further selections. 
Applying the apparent optical depth method to a subsample with $0.01 \le \tau_{\ion{H}{i}} \le 0.1$ leads to a reduced scatter (13\,\%).
For the profile-fitting procedure the scatter can be reduced as well down to about 11\,\% if only absorbers with $12.0 \le \log N_{\ion{H}{i}} \le 13.0$ are selected.
This stringent constraint minimizes the effect of systematic underestimation of $\eta$ if thermal line broadening is present.

%******************************************************************************
\section{Results and discussion}\label{results}

Having an idea of the advantages and limitations of the analysis methods, we present applications to the observed data and discuss their implications. 
In contrast to the simulated data, the observed spectra contain additional absorption features of metal lines and galactic molecular hydrogen absorption bands.
The modelling of this additional absorption, as described in Sect.\ \ref{mls}, is considered in the following analyses.
In the course of the profile-fitting analysis, the metal line parameters are fixed during the fitting procedure.
Using the apparent optical depth method, bins with a metal line or H$_2$ optical depth $\tau_{\mathrm{met}} > 0.05$ are omitted.

%------------------------------------------------------------------------------
\subsection{Profile fitting analysis}

In the \ion{He}{ii} redshift range covered by FUSE ($z \gtrsim 2.29$) we find 326 \ion{H}{i} Ly$\alpha$ lines in the Keck spectrum.
Due to the FUSE detection gap and terrestrial airglow lines, the number of \ion{H}{i} lines that are supposed to be detectable in \ion{He}{ii}, reduces to about 300. 
In addition to the \ion{He}{ii} absorbers, which are identified in \ion{H}{i}, we have to add several \ion{He}{ii} lines without any detected \ion{H}{i} counterparts.
These absorbers are about 25\,\% of all fitted \ion{He}{ii} lines.
Furthermore, we include the metal absorption lines described in Sect.\ \ref{mls}.
The resulting distribution of the column density ratio $\eta$ with redshift is shown in the lower panel of Fig.\ \ref{simple_prognose}.

Because of the high fraction of lower limits, i.e.\ \ion{He}{ii} features without \ion{H}{i} counterpart, the simple statistical mean of the absorbers with detected \ion{H}{i} and \ion{He}{ii} ($\log\eta = 1.54 \pm 1.01$, median $1.72$) is expected to be biased.
Using the Kaplan-Meier estimator \citep[e.g.][]{feigelsonnelson1985} to derive a mean value including the lower limits yields $\log\eta_{\,\mathrm{KM}} = 2.13 \pm 0.08$. 
The median of the total sample is $\log\eta = 1.93$ in good agreement with the results from the studies of the \ion{He}{ii} Ly$\alpha$ forest towards HE~2347-4342 \citep{krissetal2001, shulletal2004, zhengetal2004}, revealing $\eta\approx 80$. 

The values given above are estimated considering the total line sample.
As we argue in Sect.\ \ref{simulations}, more confident results are expected to be found if the line sample is restricted.
Therefore, we compute the mean column density ratio considering lines with $12.0 \le \log N_{\ion{H}{i}} \le 14.5$ and yield $\log\eta = 1.63 \pm 0.82$ (median 1.72).
Since for all \ion{He}{ii} features without \ion{H}{i} counterpart the lower limit of the hydrogen column density is less than $10^{12}\,\mathrm{cm}^{-2}$, no limits are included in the selected line sample.

\begin{figure}
  \centering
  \resizebox{\hsize}{!}{\includegraphics[bb=30 555 376 775,clip=]{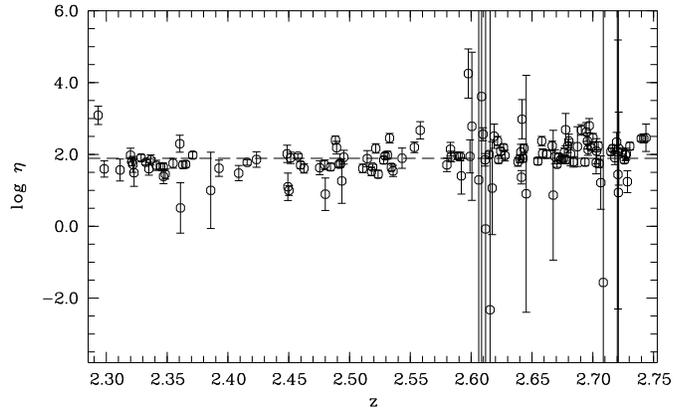}}
  \caption{Redshift distribution of the column density ratio $\eta$ for absorbers with $12.0 \le \log N_{\ion{H}{i}} \le 13.0$.
The dashed line indicates the median value $\log\eta = 1.89$.
  }
  \label{eta_z_profiles}
\end{figure}

Applying the more stringent selection criterion ($12.0 \le \log N_{\ion{H}{i}} \le 13.0$) results in $\log\eta = 1.85 \pm 0.70$ (median 1.89).
The redshift distribution of $\eta$ for this subsample is presented in Fig.\ \ref{eta_z_profiles}.
Even for the smallest sample the scatter of $\log\eta$ is roughly 40\,\%.
This is significantly more than expected by the simulations.
A possible explanation would be the redshift evolution of the column density ratio.
Due to the evolution of the sources and absorbers that compose the properties of the UV background, $\eta$ is expected to change with redshift \citep[e.g.][]{fardaletal1998}.
Our artificial spectra are generated free from any evolution.
Thus, the real data might mimic a higher overall scatter due to redshift evolution.
We estimate the column density ratio in different redshift bins of the most stringently selected sample and find indeed an increase of the mean value with redshift.
For $z < 2.50$ we yield $\log\eta \approx 1.7$, while $\log\eta \sim 2.0$ is estimated at $z > 2.60$.
However, the scatter in each redshift bin is about $20 - 40$\,\%, i.e.\ part of the scatter in the total redshift range might be due to evolution, but a significant fraction cannot be explained neither by statistical scatter nor by redshift evolution.
A Kolmogorov-Smirnov test between the results from the artificial data and the observations yields a probability of less than 1\,\% for each redshift bin that both samples are based on the same distribution function.
This leads to the conclusion that the UV background is indeed fluctuating, even though the results from the profile-fit analysis are insensitive to the scales of the $\eta$ variation \citep[an attempt to derive these scales is made by][]{fechnerreimers2005}.

\begin{figure}
  \centering
  \resizebox{\hsize}{!}{\includegraphics[bb=30 70 342 500,clip=]{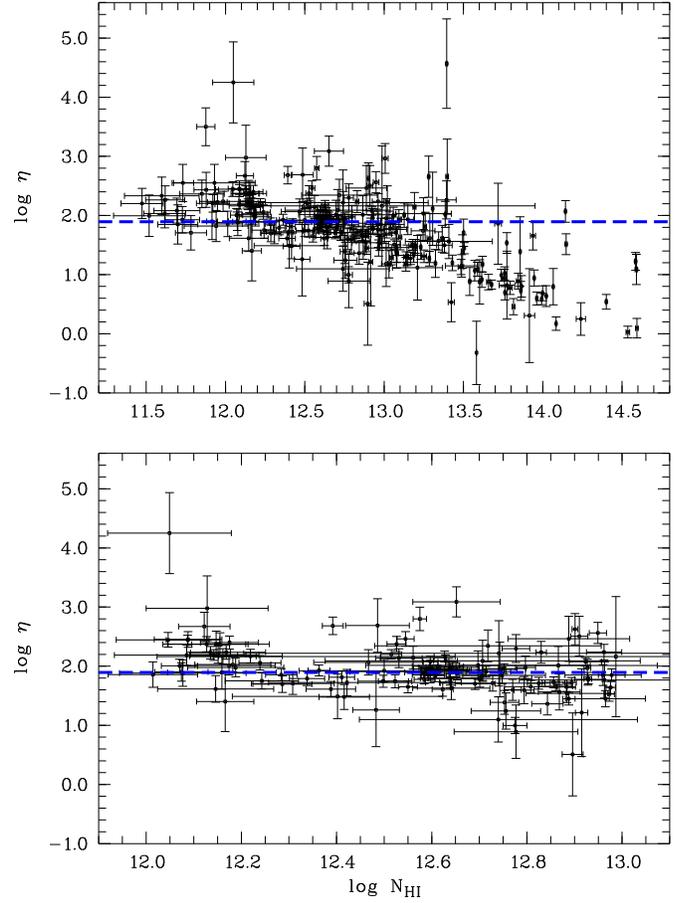}}
  \caption{Distribution of the column density ratio $\eta$ with hydrogen column density $N_{\ion{H}{i}}$ (both logarithmic). 
Only absorbers detected in \ion{H}{i} and \ion{He}{ii} with $\sigma (\log N_{\ion{H}{i},\,\ion{He}{ii}}) < 1.0$ are shown.
The upper panel presents the total sample with $\log N_{\ion{H}{i}} < 15.0$.
The lower panel is a zoom-in to $12.0 \le \log N_{\ion{H}{i}} \le 13.0$.
The dashed line represents the median value for the more constrained line sample $\log\eta = 1.89$. 
  }
  \label{HI_eta}
\end{figure}

\begin{figure*}
  \centering
  \resizebox{\hsize}{!}{\includegraphics[bb=28 595 550 778,clip=]{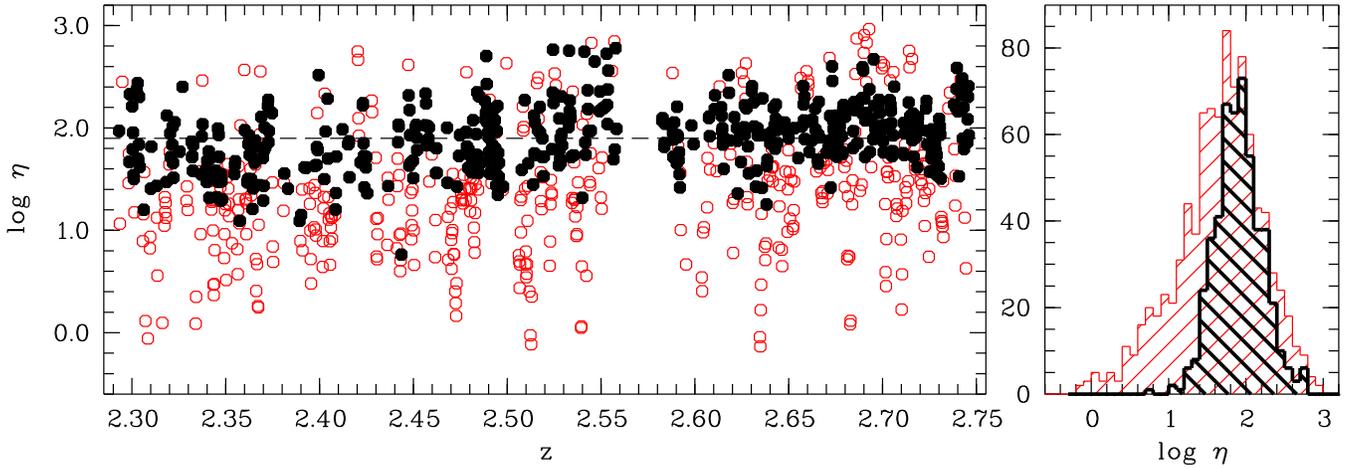}}
  \caption{Distribution of $\eta$ derived form the observed data of HS~1700+6416 using the apparent optical depth method with a bin size of $\Delta v = 20\,\mathrm{km\,s}^{-1}$ omitting bins affected by metal line absorption. 
The selected subsample ($-2.0 \le \log\tau_{\ion{H}{i}} \le -1.0$) is indicated as circles and its $\eta$ distribution is the narrower histogram in the right panel. 
The dashed line in the left panel represents the median value of the subsample $\log \eta = 1.90$.
  }
  \label{eta_observed}
\end{figure*}

For investigating the behaviour of $\eta$ with the hydrogen column density, we concentrate on absorbers with $\log N_{\ion{H}{i}} < 15.0$.
The \ion{H}{i} features stronger than this threshold are associated with the LLS in the observed redshift range.
Since LLS are believed to arise from material in the outer parts of galaxies, their column density ratio does not probe the ionizing conditions in the IGM but in a galaxy itself. 
These absorbers are excluded from the sample for the following investigation.
Furthermore, we consider only absorbers detected in both \ion{H}{i} and \ion{He}{ii} with reasonable column density uncertainties $\sigma (\log N_{\ion{H}{i},\,\ion{He}{ii}}) < 1.0$.
The distribution of $\log\eta$ with $\log N_{\ion{H}{i}}$ is shown in the upper panel of Fig.\ \ref{HI_eta}.
A clear trend is seen, that $\log\eta$ decreases with increasing \ion{H}{i} column density.
We compute the Spearman rank-order correlation coefficient yielding $r_{s} = -0.64$.
A linear fit to the data $\log\eta = a\cdot\log N_{\ion{H}{i}} + b$ leads to $a = -0.62 \pm 0.05$ and $b = 9.67 \pm 0.64$.

The analysis of the artificial data (Sect.\ \ref{simulations}) reveals that a correlation arises if a sample of lines broadened by thermal and turbulent processes is analyzed under the assumption of pure turbulent broadening.
The effect is especially noticeable for the column density ratio of absorbers stronger than $\log N(\ion{H}{i}) \gtrsim 13$, as illustrated in Fig.\ \ref{logN_logN} and discussed in the previous section \citep[see also][]{fechnerreimers2005}. 
In this case we find a Spearman rank coefficient of about $-0.25$, while no correlation is found if the assumption of pure turbulent broadening is correct ($r_{s} \approx 0$).
Indeed, the results of \citet{zhengetal2004} imply that the line width is dominated by turbulent broadening, but \citet{fechnerreimers2005} have shown that thermal broadening cannot be neglected completely.

Applying the more stringent constraint ($12.0 \le \log N_{\ion{H}{i}} \le 13.0$, presented in the lower panel of Fig.\ \ref{HI_eta}) to the observed data yields a correlation coefficient $r_{s} = -0.33$ for the observed sample, which is only half of the value for the larger sample. 
The linear fit shows now a flatter slope $a = -0.54 \pm 0.13$ and $b = 8.75 \pm 1.66$.
Also in this case, the Spearman rank coefficients indicate a slight correlation for the artificial sample with fractional thermal broadening ($r_{s} \approx -0.2)$ and no correlation for the pure turbulent broadened sample.
Since we derive a stronger anti-correlation from the observed data compared to the artificial spectra including thermal line widths, the conclusion that $\eta$ is slightly larger in voids might be correct.

%------------------------------------------------------------------------------
\subsection{Apparent optical depth method}

Considering the limitations discussed in Sect.\ \ref{simulations}, we apply the apparent optical depth method to the observed data.
In addition, we use the metal line system models of \citet{fechneretal2005a} with the modifications discussed in Sect.\ \ref{mls} to omit bins which are affected by metal line absorption in the FUSE or the Keck data.
Fig.\ \ref{eta_observed} shows the resulting $\eta$ distribution for both the whole data and the subsample selected by $-2.0 \le \log\tau_{\ion{H}{i}} \le -1.0$.
The subsample consisting of 501 points (54\,\% of the total sample) only contains values in the range of $1.0 < \log\eta < 3.0$ with a mean value of $\langle\log\eta\rangle = 1.90 \pm 0.30$ (median $1.90$).
A consideration of the total sample yields $\log\eta = 1.65 \pm 0.55$ and a median of 1.71.

Fitting a Gaussian to the $\log\eta$ distribution of the subsample yields $\log\eta_{0} = 1.903 \pm 0.013$ and $\sigma = 0.287 \pm 0.010$ corresponding to $\mathrm{FWHM} = 0.676 \pm 0.023$.
The width of the distribution is increased by more than 30\,\% in comparison to the artificial data.
Thus, we may conclude, that in addition to a scatter in $\eta$ due to noise and saturation effects a physically real variation of $\eta$ is present.  

\begin{figure}
  \centering
  \resizebox{\hsize}{!}{\includegraphics[bb=35 400 285 580,clip=]{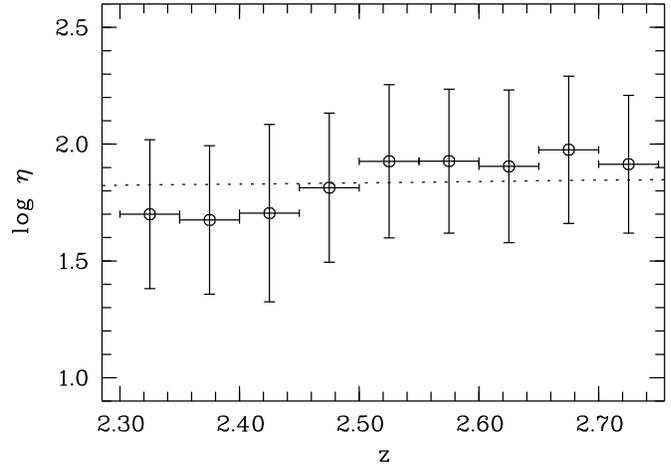}}
  \caption{Evolution of $\eta$ with redshift. 
The dotted curve represents the expected values if the ionizing sources are quasars with a spectral index of $\alpha = -1.8$ as modelled by \citet{fardaletal1998}. 
  }
  \label{eta_z}
\end{figure}

Figures \ref{eta_observed} and \ref{eta_z} show that $\eta$ is decreasing with redshift. 
The presented $\eta$ values are based on the effective optical depth of \ion{H}{i} and \ion{He}{ii}, respectively, per $\Delta z = 0.05$ bin, where censored pixels have been neglected.
In the resulting evolution $\eta$ is rising from $\log \eta = 1.70 \pm 0.32$ at $z \approx 2.33$ up to $1.98 \pm 0.32$ at $z \approx 2.68$, consistent with the results from the profile-fit analysis.
This behaviour of the \ion{He}{ii}/\ion{H}{i} ratio is expected, since it traces the evolution of the ionizing sources.
According to \citet{fardaletal1998}, the observed evolution may be mainly based on a population of quasars with spectral indices $\alpha \sim -1.8$.

An important source of error regarding the redshift evolution of $\eta$ are the uncertainties in the placement of the continuum in the FUSE spectrum, since the \ion{He}{ii} optical depth and thus the value of $\eta$ directly depends on the position of the continuum level.
This is also true for the estimate of the column density in case of line-profile fits.
As discussed in Sect.\ \ref{continuumdef} the choice of the extinction curve, which is highly uncertain in the EUV, has a strong impact on the continuum level.
Especially, if the continuum is extrapolated to a too low level at short wavelengths (e.g.\ the dotted curve in Fig.\ \ref{continuum}), the apparent optical depth and consequently the derived $\eta$ gets too small. 
Since the extinction law is wavelength-dependent, the uncertainties increase with decreasing wavelength.
Additionally, the inaccurate definition of the zero flux level at short wavelengths results in systematically underestimated \ion{He}{ii} optical depths leading to low $\eta$ values.
Thus, it might be an artifact from the continuum definition and/or zero flux level estimation that the observed evolution of $\log\eta$ appears to be steeper than predicted by the model of \citet{fardaletal1998}.
However, \citet{zhengetal2004} find a very similar evolution towards HE~2347-4342, a line of sight with a lower extinction ($E(B-V) = 0.014$).

A correlation between the \ion{H}{i} optical depth and the \ion{He}{ii}/\ion{H}{i} ratio is still present in the subsample.
The Spearman rank-order coefficient for this correlation is $r_{s} = -0.53$.
However, also the subsamples of the artificial datasets show slight correlations leading to $r_{s} \approx -0.36$ for both broadening mechanisms.
The slope describing the observed sample ($a = -0.66 \pm 0.04$) is roughly twice as steep as those from the simulated datasets with uncertainties of the same order of magnitude.
This is consistent with the results from the profile-fit analysis discussed in detail in the previous Section.
We therefore confirm the conclusion that part of the correlation between \ion{H}{i} voids and high $\eta$ values might be real.

%------------------------------------------------------------------------------
\subsection{The ionizing background}

Summarizing the results obtained by applying the apparent optical depth method and the profile-fitting procedure, we find a mean value of the column density ratio $\log\eta = 1.90$ (corresponding to $\eta \approx 80$).
The observed scatter is larger than expected from the analysis of artificial data.
Part of the enhanced scatter is possibly due to evolution of the column density ratio with redshift, which has not been incorporated in the artificial data.
Measuring $\eta$ and its scatter in smaller redshift bins suggests as well that part of the scatter must be attributed to physical effects beyond statistical noise.

\begin{figure}
  \centering
  \resizebox{\hsize}{!}{\includegraphics[bb=35 198 290 380,clip=]{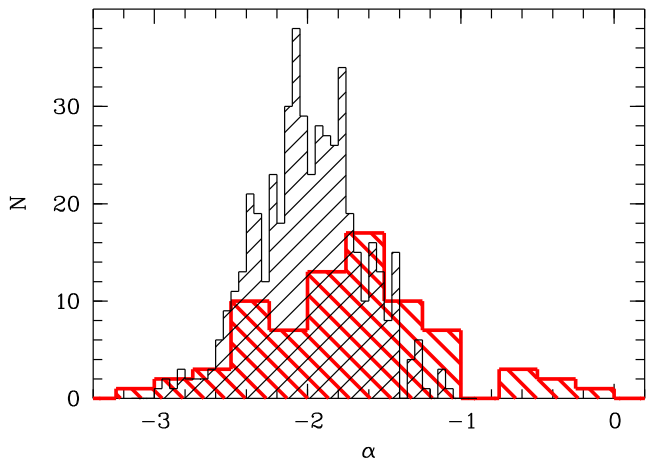}}
  \caption{Distribution of the spectral indices $\alpha$ (used as $f_{\nu} \propto \nu^{\,\alpha}$) for the observed sample (derived from the apparent optical depth result; thin line) and the whole sample from \citet[][larger bin size, thick line]{telferetal2002}.
The observed spectral indices are derived from the measured $\eta$ values using the correlation illustrated as Model A2 in Fig.\ 10 of \citet{fardaletal1998}. 
Different bin sizes are used for a better presentation.
  }
  \label{alpha}
\end{figure}

The origin of the scatter must be related to fluctuations in the \ion{He}{ii} ionizing radiation since the photoionization rate of \ion{H}{i} can be assumed to be uniform \citep[a more detailed argumentation is given by e.g.][]{boltonetal2006}.
Spatial variation at the \ion{He}{ii} ionization edge may be due to the wide range of spectral indices of the energy distribution of quasars \citep{zhengetal1997, telferetal2002, scottetal2004}.
\citet{shulletal2004} and \citet{zhengetal2004} compared the estimated column density ratios towards HE~2347-4342 to the distribution of the the spectral indices $\alpha$ of 79 quasars presented by \citet{telferetal2002}.
Both found an excess of softer radiation.
Fig.\ \ref{alpha} shows the distribution of $\alpha$ derived from the observed subsample applying the apparent optical depth method.
The $\eta$ values are converted into spectral indices using the relation between the $\ion{He}{ii}/\ion{H}{i}$ ratio and the source's spectral index presented by \citet[][their Fig.\ 10, Model A2]{fardaletal1998}.
As a mean spectral index, we find $\langle\alpha\rangle = -1.99 \pm 0.34$.
In comparison to the total sample of \citet{telferetal2002}, we confirm the excess of softer spectral indices.
One should keep in mind that the conversion from $\eta$ into $\alpha$ is based on a phenomenological model, which certainly contains additional sources of error.

Following the arguments of \citet{shulletal2004}, it is possible to convert the $\eta$ values directly into effective spectral indices.
The effective spectral index $\alpha_{\mathrm{eff}}$ describes the radiation an absorber is directly exposed to rather than the radiation once emitted by the sources, i.e.\ it denotes the spectral index of the filtered radiation of the sources.
Assuming a temperature of $T = 10^{4.3}\,\mathrm{K}$ and the equality of the local spectral indices at 1 and 4 Ryd, respectively, leads to $\eta = 1.70\cdot 4^{-\alpha_{\mathrm{eff}}}$ (for details see \citet{shulletal2004} and references therein).
Applying this relation the $\alpha$ distribution is broader and shifted to even softer indices: $\langle\alpha_{\mathrm{eff}}\rangle = -2.78 \pm 0.50$.
The results from the profile-fitting method lead to similar numbers.

\citet{boltonetal2006} have studied the observed spatial variation of the column density ratio arising from a fluctuating UV background.
According to their model, the fluctuations are based on variations in the \ion{He}{ii} photoionization rate due to properties of the emitting quasars, such as the spatial distribution, luminosity function, and the spectral energy distribution \citep[see also][]{fardaletal1998}.
\citet{boltonetal2006} find $\eta$ values scattered over $2\,\mathrm{dex}$ roughly consistent with the results from \citet{zhengetal2004}, but failed to reproduce a significant number of points with $\eta < 10$.
As discussed in the previous Sections, the results from the line profile-fitting procedure are likely to be biased due to thermal line broadening when considering high density absorbers \citep[see also][]{fechnerreimers2005}.
Although \citet{boltonetal2006} do not address this problem their Fig.\ 4 reveals an increased number of lines with a low column ratio for increasing \ion{H}{i} column density.
Considering only absorbers in a restricted column density range eliminates most of the points with very low $\eta$ (see Figs. \ref{eta_z_profiles} and \ref{eta_observed}).
We conclude that there are no absorbers exposed to radiation harder than those emitted by QSOs.
Thus, the column density ratios inferred from the data of HS~1700+6416 support the model proposed by \citet{boltonetal2006} to explain the spatial fluctuation of the column density ratio.

Apart from scatter, the average value of the column density ratio can be roughly related to the dominating sources of ionizing radiation.
The inferred value ($\eta \approx 80$) is in very good agreement with the results towards HE~2347-4342 \citep{krissetal2001, shulletal2004, zhengetal2004}.
It is consistent with the UV background of \citet{haardtmadau2001} based on QSOs only.
However, there is growing evidence that there are contributions to the ionizing background from other sources than quasars \citep[e.g.][]{heapetal2000, smetteetal2002, aguirreetal2004, boltonetal2006, kirkmanetal2005}.
The excess of soft spectral indices found from Fig.\ \ref{alpha} suggests that part of the absorbers are indeed exposed to softer radiation.

Our data reduction may still be in need of improvement concerning the zero flux level (see Sect.\ \ref{observations2}).
This is in particular true for shorter wavelengths.
In order to exclude possible biases due to the reduction problems, we estimate the mean column density ratio for wavelengths longer than $1086\,\mathrm{\AA}$ ($z > 2.58$), where the data reduction appears to be correct.
We find a median of $\log\eta = 2.01$ adopting the profile-fitting results and a slightly lower value ($\approx 1.97$) using the apparent optical depth method.
These values ($\eta \approx 93 \dots 102$) are consistent with predictions for the UV background including also young star forming galaxies \citep{haardtmadau2001} and in the same range as predicted by the models of \citet{boltonetal2006} at the same redshift based on a modified version of the UV background of \citet{madauhaardtrees1999}.

%------------------------------------------------------------------------------
\subsection{Redshift evolution of the \ion{He}{ii} optical depth}

\begin{figure}
  \centering
  \resizebox{\hsize}{!}{\includegraphics[bb=30 285 375 515,clip=]{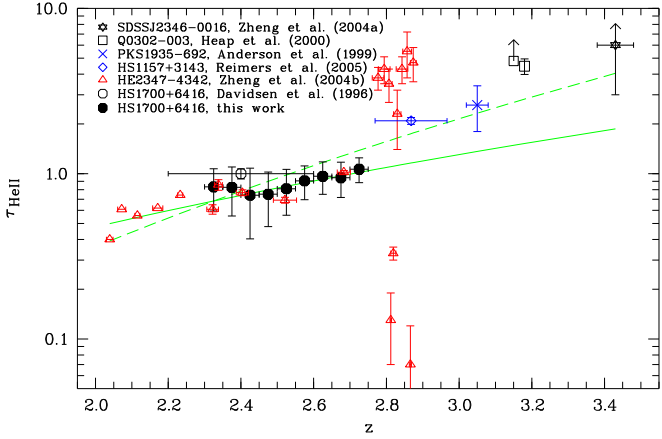}}
  \caption{Evolution of the \ion{He}{ii} opacity with redshift. 
In addition to the opacities measured in this work (filled circles), values measured towards five additional QSOs with detected \ion{He}{ii} absorption are shown.
The solid curve represents the relation $\tau_{\ion{He}{ii}} \propto (1+z)^{3.5}$ \citep{fardaletal1998} fitted to the data points $z < 2.75$. 
The dashed line indicates an overall fit to the data resulting in $\tau_{\ion{He}{ii}} \propto (1+z)^{6.2 \pm 0.4}$.
  }
  \label{tau_z}
\end{figure}

Fig.\ \ref{tau_z} shows a compilation of the measured \ion{He}{ii} opacity $\tau_{\ion{He}{ii}}$ from the literature \citep{davidsenetal1996, andersonetal1999, heapetal2000, zhengetal2004b, zhengetal2004, reimersetal2005c} together with our measurements from the FUSE spectrum of HS~1700+6416 (filled circles).
Estimating the mean normalized flux per redshift bin, pixels affected by metal line absorption are excluded.
We find a mild evolution from $\tau_{\ion{He}{ii}} = 0.83 \pm 0.24$ at $z = 2.30 - 2.35$ to $1.06 \pm 0.18$ at $z = 2.70 - 2.75$ with a slight dip ($0.74 \pm  0.34$) at $z \approx 2.45$.
These values are consistent with the opacity measurements towards HE~2347-4342 in the same redshift range \citep{zhengetal2004}.

\citet{davidsenetal1996} estimated $\tau_{\ion{He}{ii}} = 1.00 \pm 0.07$ at $\langle z \rangle = 2.4$ using a low resolution spectrum taken with the Hopkins Ultraviolet Telescope (HUT).
As can be seen from Fig.\ \ref{tau_z} our values are systematically lower by about 25\,\%.
Tests with the data uncorrected for metal line absorption suggest that this difference is partly due to the additional absorption by metal lines raising the opacity by roughly 10\,\%.
Additionally, differences in the continuum level may account for the higher effective optical depth derived by \citet{davidsenetal1996}.
However, the values are consistent on the $1\,\sigma$ level.

The solid line in Fig.\ \ref{tau_z} represents the relation $\tau_{\ion{He}{ii}} \propto (z+1)^{3.5}$ \citep{fardaletal1998} providing a suitable fit to the data at $z < 2.75$.
The value of the power law exponent $\gamma + 1$ is chosen according to the results from \ion{H}{i} Ly$\alpha$ forest statistics \citep[e.g.][]{pressetal1993, kimetal2002b}. 
Deviations from this relation at higher redshift are commonly interpreted as the evidence for the tail end of the epoch of \ion{He}{ii} reionization which completes at $z \sim 2.8$ \citep[e.g.][]{reimersetal1997, krissetal2001, zhengetal2004}.
At $z > 3$ very high values of the \ion{He}{ii} optical depth can be detected.
A simple overall fit to all data points would lead to $\tau_{\ion{He}{ii}} = 4.0\cdot 10^{-4}\,(z+1)^{6.2 \pm 0.4}$ (dashed line in Fig.\ \ref{tau_z}).
This fit should not be interpreted as a reasonable model for the evolution of $\tau_{\ion{He}{ii}}$.

%******************************************************************************
\section{Summary and conclusions}\label{conclusion}

We have presented far-UV data of the quasar HS~1700+6416 taken with FUSE, which is the second line of sight permitting us to resolve the \ion{He}{ii} Ly$\alpha$ forest.
The data are of comparable quality ($S/N \sim 5$, $R \approx 20\,000$) to those of HE~2347-4342 and cover the redshift range $2.29 \lesssim z \lesssim 2.75$.
In this redshift range, no strong variations of the \ion{He}{ii} opacity are detected.
The evolution of the effective optical depth is consistent with $(z + 1)^{3.5}$, i.e.\ this line of sight probes the post-reionization phase of \ion{He}{ii}.

The column density ratio $\eta$ has been derived using line profile fits and the apparent optical depth method, respectively.
A preliminary study of simple artificial spectra created on the basis of the statistical properties of the Ly$\alpha$ forest including or neglecting thermal line broadening, respectively, reveals the shortcomings of the standard analysis methods.
The reasons are noise, line saturation, and effects due to the assumption of pure turbulence if thermal broadening contributes to the line widths.
We find that the profile-fitting procedure leads to reliable results only for absorbers with \ion{H}{i} column densities in the range $12.0 \le \log N_{\ion{H}{i}} \le 13.0$.
The apparent optical depth method is only valid in the range $0.01 \le \tau_{\ion{H}{i}} \le 0.1$.
Otherwise, $\eta$ would be underestimated in the case of strong \ion{H}{i} absorption.
Furthermore, a scatter in $\eta$ of $10 - 15$\,\% is expected even if the underlying value is constant.

In order to avoid systematic biases due to additional absorption in the \ion{He}{ii} Ly$\alpha$ forest, a model of the metal line features expected to arise in the FUSE spectral range \citep{fechneretal2005a} has been included in the analysis.
We have also considered features of galactic H$_2$ absorption.
27\,\% of all fitted \ion{He}{ii} lines are affected by metal lines, in case of 19\,\% of the lines the derived $\eta$ value changes significantly, i.e.\ by more than $\Delta\log\eta = 0.01$.
Additionally, the required number of \ion{He}{ii} absorbers without detectable \ion{H}{i} counterpart is reduced by 30\,\% if metal lines are taken into account, and the average $\eta$ value slightly decreases.
Although the consideration of metal line absorption does not distort strongly the statistical properties of the resulting $\eta$ distribution, individual values could have been biased by up to an order of magnitude.

Photoionization models of the absorbers showing associated metal lines provide an independent estimate of $\eta$.
For these systems the \ion{He}{ii} column density can be computed based on the photoionization model for the metal line features.
Comparing the $\eta$ values directly measured from the profile-fitting procedure and inferred from the metal line system modelling leads in principle to consistent results.
A more detailed discussion will be given in a future paper \citep{fechnerreimers2005} which will also discuss the absence of the \ion{He}{ii} proximity effect towards this QSO.

For the redshift range $2.58 \lesssim z \lesssim 2.75$, where the spectrum appears to be free of artifacts due to the reduction process, the data reveals $\log\eta \approx 2.0$ on average.
This value ($\eta \approx 100$) indicates a contribution of galaxies to the UV background at these redshifts, consistent with current results from studies of the \ion{H}{i} opacity and the metallicity of the IGM.

The scatter of $\eta$ is larger than expected compared to our analysis of artificial datasets.
Therefore, we infer that the UV background might really fluctuate, even though our present results are insensitive to amplitude and scale of these variations.
The main limiting factor for a quantitative estimate of the fluctuations is the high noise level of the data.
Converting the values of the column density ratio into spectral indices, we confirm the apparent excess of soft sources as found by \citet{shulletal2004} and \citet{zhengetal2004}.
Our results suggest, that the apparent correlation between the $\eta$ value and the strength of the \ion{H}{i} absorption may be an artifact of the analysis method and does not reflect physical reality. 

Since we have shown that the assumption of pure turbulent line broadening will lead to systematical errors, the next step in the analysis of the resolved \ion{He}{ii} Ly$\alpha$ forest should be to avoid this assumption.
The apparent optical depth method applied to an appropriately restricted sample should lead to unbiased results independent of the dominant line broadening mechanism.
A bin by bin comparison as performed by the apparent optical depth method provides little potential for taking into account a potentially dominant thermal broadening, while the profile-fitting procedure could drop the assumption of pure turbulence.
However, obtaining reasonable fits would be more challenging due to the low $S/N$ of the data.
This would provide an independent strategy to estimate the IGM temperature.

\begin{acknowledgements}
This work is based on data obtained as part of FUSE guest investigator proposal C123, and on data obtained for the Guaranteed Time Team by the NASA-CNES-CSA FUSE mission operated by the Johns Hopkins University.
Financial support to U.\ S.\ participants has been provided by NASA contract NAS5-32985. 
We thank the FUSE Science and Operations team at JHU for their special efforts to schedule additional observations in the spring of 2003 and in coordination with the NASA/ESA Hubble Space Telescope. 
The HST observations were obtained at the Space Telescope Science Institute, which is operated by the Association of Universities for Research in Astronomy, Inc., under NASA contract NAS 5-26555. 
These observations are associated with program \#9982.
CF has been supported by the Verbundforschung (DLR) of the BMBF under Grant No. 50 OR 0203 and by the DFG under RE 353/49-1.
\end{acknowledgements}

\bibliographystyle{aa}
\bibliography{4950}

\end{document}